\documentclass[11pt]{article}

\usepackage{amsmath}
\usepackage{amssymb}
\usepackage{amsfonts}
\usepackage{amsthm}
\usepackage[german,english]{babel}

\usepackage{epsfig}           %
\usepackage{color}	      %

\usepackage{ifthen}
\newtheorem{theorem}{Theorem}

\newtheorem{lemma}{Lemma}

\newtheorem{corollary}{Corollary}

\newtheorem{proposition}{Proposition}

\theoremstyle{remark}
\newtheorem{remark}{Remark}

\theoremstyle{definition}

\newcommand{\intg}[4][]{{\int\limits_{#2}#3\,\mathrm{d}^{#1}#4}}
\newcommand{\Lp}[3][]{L^{#2}(#3 \ifthenelse{\equal{#1}{}}{}{,#1})}
\newcommand{\dom}[1]{\mathrm{Dom}(#1)}
\newcommand{\dm}{\mathrm{d}}
\DeclareMathOperator{\Lmoyal}{\natural}
\newcommand{\Pac}[1]{P_\mathrm{ac}(#1)}
\newcommand{\com}[3][]{[#2\, , \, #3]_{#1}}
\newcommand{\bea}{\begin{eqnarray*}}
\newcommand{\hilbert}{{\mathcal{H}}}
\newcommand{\moyal}[3][]{#2\sharp_{#1} #3}
\newcommand{\diff}[3][{}]{\frac{\mathrm{d}^{#1}#2}{\mathrm{d}{#3}^{#1}}}
\DeclareMathOperator{\support}{supp}
\newcommand{\scprd}[3][]{\langle #2\, , \, #3 \rangle_{#1}}
\newcommand{\cinfc}[1]{C^\infty_0(#1)}
\newcommand{\norm}[2][{}]{{\left\|#2\right\|}_{#1}}
\newcommand{\pdiff}[3][{}]{\frac{\partial^{#1}#2}{\partial{#3}^{#1}}}
\newcommand{\eea}{\end{eqnarray*}}
\DeclareMathOperator*{\LIM}{l.i.m.}
\newcommand{\supp}{{\rm supp\,}}
\newcommand{\mean}[1]{\langle #1\rangle}
\newcommand{\rintg}[4][]{{\int\limits_{#2}\mathrm{d}^{#1}#4\,#3}}
\newcommand{\sintg}[4]{{\int\limits_{#1}^{#2}#3\,\mathrm{d}#4}}
\newcommand{\dintg}[3]{{\int\nolimits_{#1}^\oplus #2\,\mathrm{d}#3}}
\DeclareMathOperator*{\slim}{s-lim}
\newcommand{\Set}[2]{\{#1\,|\, #2\}}
\newcommand{\nbea}{\begin{eqnarray}}
\newcommand{\acom}[3][]{\{#2\, , \, #3\}_{#1}}
\DeclareMathOperator{\ran}{Ran}
\DeclareMathOperator{\Imag}{Im}
\newcommand{\abs}[1]{{\left|#1\right|}}
\newcommand{\bounded}[2][]{\mathcal{L}^{#1}(#2)}
\newcommand{\neeq}{\end{equation}}
\newcommand{\eeq}{\end{equation*}}
\newcommand{\cinf}[1]{C^\infty(#1)}
\newcommand{\neea}{\end{eqnarray}}
\newcommand{\dform}[1]{\mathrm{d}#1}
\newcommand{\nbeq}{\begin{equation}}
\newcommand{\CN}{\mathbb{C}}
\newcommand{\beq}{\begin{equation*}}
\newcommand{\txintg}[4][]{{\int\nolimits_{#2}#3\,\mathrm{d}^{#1}#4}}
\newcommand{\txsintg}[4]{{\int\nolimits_{#1}^{#2}#3\,\mathrm{d}#4}}
\newcommand{\ointg}[4][]{{\oint\limits_{#2}#3\,\mathrm{d}^{#1}#4}}
\newcommand{\op}[2][]{\mathrm{Op}_{#1}(#2)}
\newcommand{\Order}[1]{\mathrm{O}(#1)}
\newcommand{\NN}{\mathbb{N}}
\newcommand{\RN}{\mathbb{R}}

\numberwithin{equation}{section}

\newcommand{\SAPT}{SAPT}

\newcommand{\lmoyal}[3][]{#2\natural_{#1} #3}
\renewcommand{\moyal}[3][]{\lmoyal[#1]{#2}{#3}}

\renewcommand{\dom}[1]{{\mathcal D}(#1)}
\renewcommand{\op}[2][]{\mathrm{Op}(#2)}

\title{Scattering of magnetic edge states}
\author{C. Buchendorfer, G.M. Graf\\
Theoretische Physik,
ETH-H\"onggerberg, CH--8093 Z\"urich}
\date{}
\newcommand{\D}{\mathrm{D}}

\begin{document}
\maketitle
\begin{abstract}
We consider a charged particle following the boundary of a two-dimensional
domain because a homogeneous magnetic field is applied. We develop the basic
scattering theory for the corresponding quantum mechanical edge states. The
scattering phase attains a limit for large magnetic fields which we interpret
in terms of classical trajectories.
\end{abstract}

\section{Introduction}
A charged particle moving in a domain $\Omega\subset \RN^2$ under 
the influence of a homogeneous magnetic
field $B$ may follow a skipping orbit along the boundary $\partial\Omega$. The quantum mechanical counterpart
to these orbits are extended chiral states supported near $\partial\Omega$. Under certain geometric conditions
these states give rise to some purely absolutely continuous spectrum \cite{FGW} at energies $E$ away from the Landau levels
associated with bulk states, i.e., at $E\in B\cdot \Delta$ with
\nbeq\label{Delta_condition}
\bar{\Delta}\cap (2\NN +1) =\emptyset.
\neeq
This work is about the scattering of such chiral edge states at a bent of an otherwise straight boundary $\partial\Omega$. While they,
being chiral, never backscatter, they acquire an additional phase as compared to a particle following a straight boundary of the same
length. The main result is, that this phase is proportional to the bending
angle but independent of the (large) $B$ field. We remark that the scattering
of edge states is at the basis of some theories of the quantum Hall 
effect \cite{Butt}.

The precise formulation of the setup and of the results requires some 
preliminaries. We consider a simply connected domain $\Omega\subset\RN^2$ with oriented
boundary $\partial\Omega$ consisting of a single, unbounded smooth curve $\gamma\in C^4(\RN)$ parameterized by arc length 
$s\in\RN$. We assume that $\gamma$ is eventually straight in the sense that
the curvature $\kappa(s)=\dot{\gamma}(s)\wedge\ddot\gamma(s)\in\RN$, 
($\cdot=\dform{}/\dform{s}$), is compactly supported. The bending angle
\beq
\theta:=\sintg{-\infty}{\infty}{\kappa(s)}{s}
\eeq
takes values in $[-\pi,\pi]$ and we assume
\nbeq\label{lb201}
\theta\not= \pi,
\neeq
which ensures that $\Omega$ contains a wedge of positive opening angle. 

Since the cyclotron radius, and hence the lateral extent of an edge state,
scales as $B^{-1/2}$, it will be notationally convenient to represent the 
homogeneous field as $B=\beta^2$.
The Hamiltonian is
\nbeq\label{h}
H=B^{-1}(-i\nabla- BA(x))^2=(-i\beta^{-1}\nabla-\beta A(x))^2
\neeq
on $\Lp{2}{\Omega}$ with Dirichlet boundary conditions on $\partial\Omega$. 
Here $A:\,\Omega \rightarrow \RN^2$ is a gauge field producing a unit 
magnetic field, $\partial_1A_2-\partial_2 A_1=1$. This is the usual 
magnetic Hamiltonian except for a
rescaling of energy, which is now measured in units of Landau levels spacings. 
This, or the equivalent rescaling of time, does not affect the 
scattering operator, but will simplify its analysis. 

As the dynamics of the edge states
is effectively one-dimensional, it is natural to eliminate the gauge field from its description. For the 2-dimensional system this means that we restrict
to gauges with $A_\parallel=0$ on $\partial\Omega$, i.e., 
\nbeq\label{g}
A(\gamma(s))\cdot\dot{\gamma}(s)=0.
\neeq
 
A particle moving in a half-plane $\Omega_0=\RN\times\RN_+\ni (s,u)$ will serve as a model for the asymptotic dynamics, both in the past (or at $s\rightarrow  -\infty$) and in the future (or at $s\rightarrow+\infty$). We denote the corresponding Hamiltonian on $\hilbert_0:=\Lp{2}{\Omega_0}$ by 
\nbeq\label{h0}
H_0:=(-i\beta^{-1}\partial_s+\beta u)^2+(-i\beta^{-1}\partial_u)^2,
\neeq
where we have used the Landau gauge $A=(-u,0)$.

\begin{figure}[ht]
\begin{center}
\input{domain.pstex_t}
\caption{Left: The domains $\Omega$, $\Omega^e$,
  $\Omega_\pm^e$.\label{fig3}} 
\caption{Right: The domains $\Omega_0$, $\Omega^e_0$, $\Omega^e_{0\pm}$.\label{fig4}}
\end{center}
\end{figure}

To serve as scattering asymptotes, states in $\Lp{2}{\Omega_0}$ have to 
be identified with states in $\Lp{2}{\Omega}$.
To this end we introduce the tubular map:
\begin{gather}
\mathcal{T}:\, \Omega_0\, \rightarrow\,\RN^2\nonumber\\
x(s,u)\equiv \mathcal{T}(s,u)=\gamma(s)+u\varepsilon\dot{\gamma}(s),\label{tubmap}
\end{gather}
where $\varepsilon=\left(\begin{array}{cc}0&-1\\1&0\end{array}\right)$ is the rotation by $\pi/2$, and hence $\varepsilon\dot{\gamma}(s)$ the
inward normal. The map $\mathcal{T}$ is injective on 
\beq
\Omega_0^e:=\Set{(s,u)\in\Omega_0}{s\in\RN,\, 0\leq u <w(s) },
\eeq
with Jacobian $|\det\D \mathcal{T}|=1-u\kappa(s)$ uniformly bounded away from zero, 
for some sufficiently small positive, continuous width function $w(s)$. 
Due to condition \eqref{lb201} we may take it so that 
\nbeq\label{w}
w(s)\geq c_1+c_2\abs{s}
\neeq
for some $c_1,\,c_2>0$. The map \eqref{tubmap} provides coordinates 
$(s,u)$ on the image $\Omega^{e}:=\mathcal{T}(\Omega^{e}_{0})\subset
\Omega$ (Fig. \ref{fig3}, \ref{fig4}). 
Not all of 
$\Omega_{0}^e$ is essential for the sought identification, but only its tails 
near $s=\pm\infty$,  
\beq
\Omega_{0\pm}^e:=\Set{(s,u)\in\Omega_{0}^e}{\pm s>C }.
\eeq
For large enough $C$ the tubular map is Euclidean if restricted to
$\Omega_{0\pm}^e$, since $\supp\ddot{\gamma}$ is compact. To make the dynamics
of \eqref{h} and \eqref{h0} comparable, we assume that 
\nbeq\label{g1}
A(x)=(-u,0),\qquad (x\in \Omega_\pm^e)
\neeq
w.r.t. the Euclidean coordinates $(s,u)$ in 
$\Omega_\pm^e:=\mathcal{T}(\Omega_{0\pm}^e)$. This does not fix the potential 
$A$ outside of $\Omega_-^e\cup\Omega_+^e$ beyond the condition \eqref{g}.
Any residual gauge transformation $A\rightarrow A+\nabla\chi$ in $\Omega$ 
consistent with these requirements has $\chi(x)$ constant in 
$\Omega_-^e\cup\Omega_+^e$. In fact, $\chi(x)$ takes constant values $\chi_\pm$
separately on $\Omega_\pm^e$, and
\nbeq\label{chi}
\chi_+-\chi_-=
\sintg{-\infty}{\infty}{\nabla\chi(\gamma(s))\cdot\dot{\gamma}(s)}{s}=0.
\neeq

The asymptotic Hilbert space $\Lp{2}{\Omega_0}$ is now mapped into
$\Lp{2}{\Omega}$ by means of 
\begin{gather}
J:\, \Lp{2}{\Omega_0} \,\rightarrow \,\Lp{2}{\Omega}\nonumber\\
(J\psi)(x)=\begin{cases}
j(u-w(s))\psi(s,u),&\text{ if } x=x(s,u)\in \Omega^{e},\\
\quad 0,& \text{ otherwise, }
\end{cases}\label{lb120}
\end{gather}
where $j\in\cinf{\RN}$, $j\leq 1$ is such that 
\nbeq \label{lb121}
j(u)=\begin{cases}
        1,& u\leq -2w_0,\\
        0,& u\ge -w_0,
\end{cases}
\neeq
for some $w_0$. The purpose of the transition function $j$ is to make $J\psi$
as smooth as $\psi$. If $w_0$ is large enough, 
$\supp J\psi\subset \Omega_-^e\cup\Omega_+^e$; if, on the other hand, $w_0$ is
small enough, we have $J\psi(x)=\psi(s,0)$ for all $x=x(s,0)\in\partial\Omega$.

The first result establishes the usual properties of scattering.

\begin{theorem}\label{theorem}
The wave operators
\begin{gather*}
W_\pm:\,\Lp{2}{\Omega_0}\,\rightarrow \Lp{2}{\Omega}\\
W_\pm:=\slim_{t\to\pm\infty}e^{iHt}Je^{-iH_0t}
\end{gather*}
exist and are complete:
\beq
\ran W_\pm=\Pac{H}\hilbert.
\eeq
Moreover, $W_\pm$ are isometries and do not depend on the choice of $w,\,j$ in
the definition of $J$. \end{theorem}

\begin{remark}
Under a residual gauge transformation the wave operators transform as 
\beq
W_\pm\rightarrow e^{i(\chi_\pm-\chi(x))}W_\pm,
\eeq
implying by \eqref{chi} that the scattering operator $W^*_+W_-$ is invariant.
\end{remark}

We next consider the limit where $\beta$ grows large while the energy, 
rescaled as
in \eqref{h}, is kept fixed. The limit of the scattering operator is thus 
best formulated in a scheme where edge states with fixed energy are displayed
as being independent of $\beta$. The domain $\Omega_0$ is invariant under 
scaling 
\nbeq\label{scal}
x\to \beta x
\neeq
and the Hamiltonian transforms as
\nbeq\label{h0t}
H_0\cong -\partial_u^2+(-i\partial_s+u)^2,
\neeq
which shows that the spectrum of $H_0$ is independent of $\beta$. Let 
$\hilbert_T:=\Lp{2}{\RN_+,\mathrm{d}u}$ be the space of transverse wave
functions, on which $-\partial_u^2$ acts with a Dirichlet boundary 
conditions at $u=0$. The translation invariance in $s$ of (\ref{h0t}) calls for
the (inverse) Fourier transform
\begin{gather}
\mathcal{F}_\beta:\, \dintg{}{\hilbert_T}{k}\cong\Lp{2}{\RN,\hilbert_T} 
\,\rightarrow\,\Lp{2}{\Omega_0},\qquad
\psi=\dintg{}{\psi(k)}{k}\,\mapsto\,\mathcal{F}_\beta\psi,\nonumber\\
(\mathcal{F}_\beta\psi)(k)=
\LIM_{K\to\infty}\frac{\beta^{1/2}}{(2\pi)^{1/2}}
\sintg{-K}{K}{e^{i\beta ks} \mathcal{D}_\beta\psi(k)}{k},\label{ft}
\end{gather}
where the scaling of $x=(s,u)$ has been incorporated for $u$ by means of
\nbeq\label{dbeta}
 \mathcal{D}_\beta:\, \hilbert_T\,\rightarrow\,\hilbert_T,
\qquad
( \mathcal{D}_\beta\psi)(u)=\beta^{1/2}\psi(\beta u),
\neeq
and for $s$ explicitly in the integral. (It is, in a precise sense,
a Bochner integral of $\hilbert_T$-valued functions \cite[Sec. 1.1,Sec. 1.8]{ABHN}).
Then
\nbeq\label{h0k}
\mathcal{F}_\beta^{-1}H_0\mathcal{F}_\beta=\widehat{H}_0:=
\dintg{}{H_0(k)}{k},\qquad H_0(k)=-\partial_u^2+(k+u)^2.
\neeq
The fiber $H_0(k)$, see \cite{BP}, has simple, discrete 
spectrum $\{E_n(k)\}_{n\in\NN}$ with projections denoted as $P_n(k)$. 
The energy curve $E_n(k)$, called
the $n$-th deformed Landau level, is a smooth function of $k$
increasing from $2n+1$ to $+\infty$ for $k\in(-\infty,\infty)$ with 
$E'_n(k)>0$.
The corresponding normalized eigenvectors by $\psi_n(k)$ may be taken 
as smooth functions (in $\hilbert_T$-norm) of $k$, though the choice
is affected by the arbitrariness of their phase, 
\nbeq\label{ph}
\psi_n(k)\mapsto\ e^{i\lambda_n(k)}\psi_n(k).
\neeq
They decay exponentially in $u$ (see Lemma~\ref{lb112}). 

In this scheme the scattering operator is
\nbeq\label{so}
S=\mathcal{F}_\beta^{-1}W^*_+W_-\mathcal{F}_\beta\,:
\dintg{}{\hilbert_T}{k}\,\rightarrow\,\dintg{}{\hilbert_T}{k},
\neeq
and becomes independent of the magnetic field if large:

\begin{theorem}\label{theorem2} We have 
\nbeq\label{sl}
\slim_{\beta\to\infty}S=S_\phi
\neeq 
with 
\begin{gather}
S_\phi=\dintg{}{\sum_{n}e^{i\phi_n(k)}P_n(k)}{k},\nonumber\\
\phi_n(k)=-\frac{E_n^{(1)}(k)}{E_n'(k)}\theta,\label{phi}\\
E^{(1)}_n(k)=\scprd{\psi_n(k)}{H_1(k)\psi_n(k)},\nonumber\\
H_1(k)=u^3+3u^2k+2uk^2.\nonumber
\end{gather}
More precisely, if energies are restricted to any open interval $\Delta$ 
between Landau levels, as in (\ref{Delta_condition}), 
the limit (\ref{sl}) holds in norm: for any $\varepsilon>0$ there is 
$C_{\Delta,\varepsilon}$ such that
\nbeq\label{se}
\norm{(S-S_\phi)E_{\Delta}(\widehat{H}_0)}\leq C_{\Delta,\varepsilon} 
\beta^{-1+\varepsilon}.
\neeq
\end{theorem}

\begin{remark}
$E_n^{(1)}(k)$ is the first order correction to the eigenvalue 
$E_n(k)$ under the (singular) perturbation $\beta^{-1}\kappa(s)H_1(k)$ 
of $H_0(k)$ due to the curvature of the boundary.
\end{remark}

We conclude with some comments about the origin of the phase $\phi_n$. The 
Hamiltonian \eqref{h} results from the quantization of mixed systems \cite{LF}
in the sense that it may be regarded as the quantization over the phase 
space $\RN^2\ni(s,k)$ of the classical symbol
\nbeq\label{sym}
H(s,k)=H_0(k)+\beta^{-1}\kappa(s)H_1(k),
\neeq
which formally takes values in the operators on $\hilbert_T$. Typical WKB
solutions for such systems have a phase consisting of a dynamical part of
$\Order{\beta}$ followed by a geometric part, namely 
the Berry and Rammal-Wilkinson phases, 
$\gamma_\mathrm{B}+\gamma_\mathrm{RW}$, which are of $\Order{1}$.
The scattering operator $S$ discounts from this the phase that pertains to the
principal symbol $H_0(k)$ alone. The phase left over thus stems from the
sub-principal symbol only, with the two parts now suppressed by a factor
$\beta^{-1}$. The phase \eqref{phi} is thus dynamical --- despite its
connection with the geometry of $\Omega$ ---, while for the geometric ones 
we find $\beta^{-1}$ times
\begin{gather}
\gamma_\mathrm{B}(s,k)=\kappa(s)\frac{E_n^{(1)}(k)}{E_n'(k)}
\Imag{\scprd{\psi_n(k)}{\partial_k\psi_n(k)}},\label{lb705}\\
\gamma_\mathrm{RW}(s,k)=-\gamma_\mathrm{B}(s,k)+\frac{\kappa(s)}{E_n'(k)}
\Imag{\scprd{H_1(k)\psi_n(k)}{\partial_k\psi_n(k)}}.\label{lb706}
\end{gather}
In the next section we give a heuristic interpretation of the
edge states and of the scattering phase $\phi_n(k)$ in terms
of classical orbits bouncing at the boundary. Related considerations are found
in \cite{HSm}. Readers more interested in the
proofs may proceed without loss to Sects.~\ref{proofs},
\ref{proofs2}. Higher order corrections like (\ref{lb705}, \ref{lb706}) are
discussed in Sect. \ref{hoa}.

\section{Classical trajectories and scattering phase}
\label{classtraj}

The Hamiltonian 
\beq
H_0=(\beta^{-1}p_s+\beta u)^2+\beta^{-2}p_u^2,
\eeq 
which is the classical counterpart to \eqref{h0}, has circular trajectories 
for which radius $r>0$ and velocity $v\in\RN^2$ are in the fixed relation 
$r=|v|/2$. 
Some of them bounce along the edge of the half-plane. Their shape
may be parameterized in various ways: (i) By the ratio 
\nbeq\label{sc2}
\frac{k}{r}=\cos{\eta}
\neeq
between the distance $k=\beta^{-2}p_s$ of the guiding center to 
$\partial\Omega_0$ (negative, if inside $\Omega_0$) and the radius $r$. This 
is also expressed through the angle $\eta$ between the boundary and the arc, 
see Fig.~\ref{fig1}. (ii) By the
ratio 
\nbeq\label{sc3}
\frac{v_\parallel}{|v|}=\frac{\sin{\eta}}{\eta}
\neeq
between the average velocity $v_\parallel$ along the edge and the (constant)
velocity $|v|$ or, equivalently, between the length $2r\sin{\eta}$ of the chord 
and $2r\eta$ of the arc in Fig.~\ref{fig1}. 

\begin{figure}[ht]

\begin{center}
\input{fig.pstex_t}
\caption{Left: a bouncing trajectory}\label{fig1} 
\caption{Right: the phase space of transversal motion} \label{fig2}
\end{center}
\end{figure}

We now turn to the quantum state $e^{iks}\psi_n(k)$ for $\beta=1$,
cf.~\eqref{ft}. On the basis of \eqref{sc2} it may be associated, at least
asymptotically for large $n$, with a classical trajectory of shape $\eta$ if 
\nbeq\label{sc4}
k_n=\sqrt{E_n(k_n)}\cos{\eta}.
\neeq
The same conclusion is reached on the basis of
\eqref{sc3} if $v_\parallel$ is identified with the group velocity $E'_n(k)$,
as we presently explain.
The phase space $\RN_+\times\RN\ni(u,p_u)$ underlying 
$H_0(k)$ is shown in Fig.~\ref{fig2}, together with a trajectory of energy 
$(k+u)^2+p_u^2=E$. Let $A(E,k)$ be the area of the cap inside this trajectory.
The Bohr-Sommerfeld condition, whose asymptotic validity we take for 
granted, states that
$A(E_n(k),k)=2\pi n$, ($n\in\NN$), and derivation w.r.t. $k$ yields 
\beq%
\frac{\partial A}{\partial E}E'_n(k)+\frac{\partial A}{\partial k}=0.
\eeq
Using that $-\partial A/\partial k$ is the length of the chord in 
Fig.~\ref{fig2}, and $\partial A/\partial\sqrt{E}=
2\sqrt{E}(\partial A/\partial E)$ that of the arc we find 
\nbeq\label{sc6}
\frac{v_\parallel}{|v|}=\frac{E'_n(k)}{2\sqrt{E_n(k)}}=
\frac{\sin{\eta}}{\eta},
\neeq
provided $k=k_n$ is chosen as in \eqref{sc4}. The energy is then
$E_n(k_n)\propto n$ and the radius before the scaling \eqref{scal} is given
by $r_n^2=\beta^{-2}E_n(k_n)$. 

In light of this correspondence we shall 
discuss the motion along a curved boundary. The semiclassical limit, 
$n\gg 1$, and the limit
of small curvature, $\kappa(s)r_n\ll 1$, are consistent as long as 
$1\ll n \ll \beta^2\kappa^{-2}$, i.e., for large magnetic fields. 
We again first deal with the classical particle, whose incidence 
angle $\eta$ may now slightly change from hit to hit. Let 
\beq%
G(s,s';\kappa)=\intg{\gamma}{p\cdot}{x}
\eeq
be the (reduced) action along one of the two arcs $\gamma$ of radius $r$
joining neighboring collision points $s$ and $s'$ along the boundary 
of curvature $\kappa(\cdot)$ (provided they are close enough, so that the arcs
exist). With $p=\beta^2(v/2+A)$ we obtain \cite{BK}
\beq%
G(s,s';\kappa)=\beta^2(r\mathcal L-\mathcal A),
\eeq
where $\mathcal L$ is the length of the arc $\gamma$ and $\mathcal A$ the area
between the arc and the boundary $\partial\Omega$. In fact,
\beq%
\frac{1}{2}\intg{\gamma}{v\cdot}{x}=
\frac{|v|}{2}\intg{\gamma}{\diff{x}{\sigma}\cdot}{x}=r\intg{\gamma}{}{\sigma},
\eeq
where $\sigma$ is the arc length along $\gamma$; and, by Stokes' theorem,
$\intg{\gamma}{A\cdot}{x}=-\mathcal A$, because the arc is traversed clockwise
and because of \eqref{g}. We next consider an arc starting at $s$ with angle
$\eta$ and look for the dependence of $s'-s,\,\eta'-\eta$ and $G(s,s';\kappa)$
up to first order in a small curvature $\kappa$. Elementary considerations 
show that 
\begin{gather*}%
\delta(s'-s)\approx -\kappa r^2\sin{2\eta},\qquad
\delta(\eta'-\eta)\approx 0,\\
\delta\mathcal L\approx -2\kappa r^2\sin{\eta},\qquad
\delta\mathcal A\approx -\frac{4}{3}\kappa r^3\sin^3{\eta},\\
\delta G(s,s';\kappa)\approx 
-2\beta^2r^3\kappa\sin{\eta}\bigl(1-\frac{2}{3}\sin^2{\eta}\bigr),
\end{gather*}
where $\kappa=\kappa(\tilde s)$ for any $\tilde s$ between $s$ and $s'$. 
We then take a number $m$ of hops $s_i= s_i[\kappa],\,(i=0,\ldots m)$
sufficient to cover the bent $\supp\kappa$. Using $s_i-s_{i-1}=2r\sin{\eta}$
for $\kappa\equiv 0$, we compute in the small curvature limit
\begin{gather*}%
\delta(s_m-s_0)=\sum_{i=1}^m\frac{\delta(s_i-s_{i-1})}{s_i-s_{i-1}}
(s_i-s_{i-1})\approx 
-r\frac{\sin{2\eta}}{2\sin{\eta}}\sintg{-\infty}{\infty}{\kappa(s)}{s}
=-r\theta\cos{\eta},\\
\delta\sum_{i=1}^m G(s_{i-1},s_i;\kappa)
\approx-\beta^2 r^2\theta\bigl(1-\frac{2}{3}\sin^2{\eta}\bigr).
\end{gather*}
An incoming quantum wave $e^{i\beta ks} \mathcal{D}_\beta\psi(k)$ 
should therefore gather an additional phase 
\beq%
\phi_n(k)=-\beta k\delta(s_m-s_0)+\delta\sum_{i=1}^m G(s_{i-1},s_i;\kappa)
\eeq
as compared to one following a straight boundary of the same length. With 
\eqref{sc4} we find
\nbeq\label{sc13}
\phi_n(k_n)=\beta r_n\theta\sqrt{E_n(k_n)}\cos^2{\eta}
-\beta^2 r_n^2\theta\bigl(1-\frac{2}{3}\sin^2{\eta}\bigr)
=-\frac{1}{3}\theta E_n(k_n)\sin^2{\eta}.
\neeq
On the other hand, the phase $\phi_n(k_n)$ may be computed from
\eqref{phi}. Since the trajectory in Fig.~\ref{fig2} is traversed at a uniform
rate, expectations w.r.t. $\psi_n(k_n)$ reduce in the limit to integrations
w.r.t. $(2\eta)^{-1}\dm\alpha$, where a point on the arc is represented by its
angle $\alpha\in[-\eta,\eta]$ as seen from the center of the circle. We
rewrite $\sqrt{E}\cos{\alpha}=k+u=:u'$ and $u^3+3u^2k+2uk^2=u'(u'^2-k^2)$,
use 
\beq%
\frac{1}{2\eta}\sintg{-\eta}{\eta}{\cos{\alpha}}{\alpha}
=\frac{\sin{\eta}}{\eta},\qquad
\frac{1}{2\eta}\sintg{-\eta}{\eta}{\cos^3{\alpha}}{\alpha}
=\frac{\sin{\eta}}{\eta}\bigl(1-\frac{1}{3}\sin^2{\eta}\bigr),
\eeq
and obtain
\begin{gather*}%
E^{(1)}_n(k)\approx
E_n(k_n)^{3/2}\frac{\sin{\eta}}{\eta}
\bigl(1-\frac{1}{3}\sin^2{\eta}-\cos^2{\eta}\bigr)
=\frac{2}{3}E_n(k_n)^{3/2}\frac{\sin^3{\eta}}{\eta},\\
\phi_n(k_n)\approx-\frac{1}{3}\theta E_n(k_n)\sin^2{\eta},
\end{gather*}
where we used \eqref{sc6} in the last step. The result is in agreement 
with \eqref{sc13}.

\section{Existence and completeness of wave operators}\label{proofs}
Existence and completeness of the wave operators $W_\pm$ follow in a
rather standard way from propagation estimates for the dynamics $e^{-iHt}$ and
$e^{-iH_0t}$. 

Such an estimate is established in the second part of the 
following lemma. It depends on a Mourre estimate \cite{FGW}, which in 
turn rests on a geometric property discussed in the first part: 

\begin{lemma}\label{propagationestimate}
\begin{enumerate}
\item There is a function $\sigma\in C^2(\bar{\Omega})$ extending arc length from $\partial\Omega$ to $\Omega$, i.e., $\sigma(\gamma(s))=s$ for 
$s\in\RN$, satisfying
\nbeq\label{extofarclength}
\norm[\infty]{\partial_i\sigma}<\infty,\quad \norm[\infty]{\partial_i\partial_j\sigma}<\infty.
\neeq

\item For any $\varepsilon>0$, $\alpha>1/2$ and $\Delta$ as in \eqref{Delta_condition}:
\nbeq\label{propestm}
\sintg{-\infty}{\infty}{\norm{\mean{\sigma}^{-\alpha}e^{-iHt}E_{\Delta}(H)\psi}^2}{t}\leq C_{\Delta,\alpha} \beta^{1+\varepsilon}\norm{\psi}^2
\neeq
with $C_{\Delta,\alpha}$ independent of large enough $\beta$.
\end{enumerate}
\end{lemma}

\begin{proof}
1. On $\Omega_0$ we may choose the following extension of arc length:
\nbeq\label{s0}
\sigma_0(s,u):=\frac{s}{w(s)}(w(s)-u)j(u-w(s)).
\neeq 
It satisfies \eqref{extofarclength} and is supported on $\Omega^{e}_{0}$.  We therefore obtain an extension of arc length $\sigma(x)$ from $\partial\Omega$ to  $\Omega$  by transforming $\sigma_0$ under the tubular map:
\beq
\sigma(x):=\begin{cases} \sigma_0(s,u) & 
\text{ if } x=x(s,u)\in \Omega^{e},\\
        0& \text{ otherwise.}
\end{cases}
\eeq 
$\sigma$ satisfies \eqref{extofarclength} because $\sigma_0$ is an extension of arc length on $\Omega_0$, $\sigma$ is supported on $\Omega^{e}$
  and the inverse tubular map has bounded first and second derivatives on $\Omega^e$. The extension of $\sigma$ by zero to the complement of $\Omega^{e}$ is smooth by construction of $j$.

2. To better display the dependence on $\beta$ 
of some of the bounds below we scale $\Omega$ to 
$\tilde{\Omega}=\beta\Omega$, so that 
$H\cong \tilde{H}$, where
\beq
\tilde{H}=(-i\nabla-\tilde{A})^2,
\eeq
on $\Lp{2}{\tilde{\Omega}}$ with $\tilde{A}(x)=\beta A(x/\beta)$ 
corresponding to a unit magnetic field. The corresponding extension of 
arc length from part (1) is $\tilde{\sigma}(x)=\beta\sigma(x/\beta)$. 
We claim that for given $E\not\in 2\NN+1$
\begin{gather}
\norm{\com{\tilde{H}}{\tilde{\sigma}}(\tilde{H}+i)^{-1}}\leq C,\label{l1}\\
\norm{\com{\com{\tilde{H}}{\tilde{\sigma}}}{\tilde{\sigma}}}\leq C,\label{l2}\\
E_{\tilde{\Delta}}(\tilde{H}) i\com{\tilde{H}}{\tilde{\sigma}}E_{\tilde{\Delta}}(\tilde{H})\geq c E_{\tilde{\Delta}}(\tilde{H})\label{l3}
\end{gather}
with $C,c >0$ and an open interval $\tilde{\Delta}\ni E$, all independent of 
$\beta$ large. Indeed, (\ref{l1}, \ref{l2}) follow from
\begin{gather*}
i\com{\tilde{H}}{\tilde{\sigma}}=(-i\nabla-\tilde{A})\cdot\nabla\tilde{\sigma}+\nabla\tilde{\sigma}\cdot(-i\nabla-\tilde{A}),\\
i\com{i\com{\tilde{H}}{\tilde{\sigma}}}{\tilde{\sigma}}=2(\nabla\tilde{\sigma})^2,
\end{gather*}
and \eqref{l3} has been shown in connection with the proof of Thm. 3 in 
\cite{FGW}. The bounds (\ref{l1}-\ref{l3}) now imply \cite{HS} for 
$\alpha>1/2$:
\beq
\sintg{-\infty}{\infty}{\norm{\mean{\tilde{\sigma}}^{-\alpha}e^{-i\tilde{H}t}E_{\tilde{\Delta}}(\tilde{H})\psi}^2}{t}\leq C \norm{\psi}^2.
\eeq
Undoing the unitary scale transformation, this amounts to:
\beq
\beta^{-2\alpha}\sintg{-\infty}{\infty}{\norm{(\sigma^2+\beta^{-2})^{-\alpha/2}e^{-iHt}E_{\tilde{\Delta}}(H)\psi}^2}{t}\leq C\norm{\psi}^2.
\eeq
Using a covering argument for $\Delta$, this proves
\nbeq\label{lb206}
\sintg{-\infty}{\infty}{\norm{\mean{\sigma}^{-\alpha}e^{-iHt}E_{\Delta}(H)\psi}^2}{t}\leq C_\alpha \beta^{2\alpha}\norm{\psi}^2,
\neeq
for $\beta\geq 1$, which may be assumed without loss. For 
$\alpha \leq (1+\varepsilon)/2$ the claim follows from $\beta^{2\alpha}\leq
\beta^{1+\varepsilon}$. It then extends to $\alpha > (1+\varepsilon)/2$ 
because the l.h.s of \eqref{lb206} is decreasing in $\alpha$. 
\end{proof}

\begin{remark}
The bound \eqref{propestm} may be understood in simple terms. The velocity of a
particle tangential to the boundary is 
$i[H,s]= \beta^{-1}(-i\beta^{-1}\nabla-\beta A(x))\cdot\nabla
s=O(\beta^{-1})$, assuming its energy $H$ lies in $\Delta$. It therefore
takes the particle a time $O(\beta)$ to traverse a fixed piece of 
the boundary such as the bent. Eq.~\eqref{propestm} is stating just this, up 
to a multiplicative error $O(\beta^{\varepsilon})$.
\end{remark}

We shall prove existence and completeness of the wave operators
$W_\pm$ by local Kato smoothness. More precisely by
\cite[Thm. XIII.31]{RS3} or, with more detail, by 
\cite[Sect. 4.5, Thm. 1, Cor. 2, Rem. 3, Thm. 6]{Y} all of Thm.~\ref{theorem},
except for the uniqueness statement, is implied by the following lemma:

\begin{lemma}
\begin{enumerate}
\item 
$J$ maps $\dom{H_0}$ into $\dom{H}$. Moreover
\nbeq \label{l4}
HJ-JH_0=\sum_{i=1}^{2} A^*_iM_iA_i^0,
\neeq
where $A_i^{(0)}$ are $H_{(0)}$-bounded and $H_{(0)}$-smooth on $\Delta$, and $M_i$ are\\ bounded operators, ($i=1,2$, $(0)=0$ or its omission).
\item 
\nbeq\label{l4a}
\slim_{t\to\pm\infty}(1-JJ^*)e^{-iHt}E_{\Delta}(H)=0.
\neeq
\end{enumerate}
\end{lemma}

\begin{proof}

1. For $C$ large enough, $|\sigma_0(s,u)|>C$ implies $j(u-w(s))=1$. In fact,
if $j(u-w(s))<1$ we have $u-w(s)>-2w_0$ and therefore, see eq.~\eqref{s0}, 
\nbeq\label{l4b}
|\sigma_0(s,u)|=\frac{|s|}{w(s)}(w(s)-u)j(u-w(s))
\neeq
is bounded by $2w_0\sup_s|s|/w(s)$, which is finite by \eqref{w}. By 
\eqref{l4b} we also see that $|\sigma_0(s,u)|>C$ implies that $|s|$ is large. 
These two implications, together with \eqref{g1}, show that 
$(HJ-JH_0)F(|\sigma_0|>C)=0$, where $F(x\in A)$ is the characteristic
function of the set $A$. Together with a similar relation for $\sigma$
instead of $\sigma_0$ we obtain 
\nbeq \label{l5}
HJ-JH_0=\chi(HJ-JH_0)\chi_0,
\neeq
where $\chi_{(0)}=F(|\sigma_{(0)}|\le C)$.

Eq.~\eqref{l5} may be written in the form \eqref{l4} with
\begin{align*}
A_1&=\mean{\sigma}^{-\alpha}(H-i),\\
M_1&=\mean{\sigma}^{\alpha}(H+i)^{-1}\chi H J \chi_0\mean{\sigma_0}^{\alpha},\\
A_1^0&=\mean{\sigma_0}^{-\alpha},\\
A_2&=\mean{\sigma}^{-\alpha},\\
M_2&=-\mean{\sigma}^{\alpha} \chi J H_0 \chi_0 (H_0+i)^{-1}\mean{\sigma_0}^{\alpha},\\
A_2^0&=\mean{\sigma_0}^{-\alpha}(H_0+i).
\end{align*}
The claimed properties about the $A_i^{(0)}$ hold true by \eqref{propestm} and we are left to show those of the $M_i^{(0)}$. Since $\chi\mean{\sigma}^{\alpha}$, $\chi_0\mean{\sigma_0}^{\alpha}$
(and $J$) are bounded, we need to show that 
\begin{multline*}
H\chi(H+i)^{-1}\mean{\sigma}^{\alpha}\\
=H\chi [\mean{\sigma}^{\alpha}(H+i)^{-1}+(H+i)^{-1}\com{\mean{\sigma}^{\alpha}}{H}(H+i)^{-1}]
\end{multline*}
is, too (and similarly for the '$0$'-version). Indeed, for $\alpha<1$, $\com{\mean{\sigma}^{\alpha}}{H}(H+i)^{-1}$ is bounded, cf.~\eqref{l1}, and so is 
\beq
Hf(H+i)^{-1}=H(H+i)^{-1}\bigl(f+\com{H}{f}(H+i)^{-1}\bigr)
\eeq
for $f=\chi\mean{\sigma}^{\alpha}$ or $f=\chi$.

2. Since $(1-JJ^{*})(1-\chi)=0$ and $\chi\mean{\sigma}^{\alpha}$ is bounded, 
we may show
\beq
\lim_{t\to\pm\infty}\mean{\sigma}^{-\alpha}e^{-iHt}E_{\Delta}(H)\psi=0.
\eeq
As a function of $t$, this state has bounded derivative and is square 
integrable in $t$, cf.~\eqref{propestm}. Hence the claim.
\end{proof}

It remains to show that $W_\pm=W_\pm(J)$ is independent of $j$
and $w$ in the construction \eqref{lb120} of $J$. We may choose 
$\tilde\jmath,\,\tilde w$ still satisfying the requirements 
(\ref{lb121}, \ref{w}) and, moreover, 
\begin{gather}
\supp \tilde\jmath(u-\tilde w(s))\subset\Omega_-^e\cup\Omega_+^e,\label{jt1}\\
\tilde\jmath(u-\tilde w(s)) j(u-w(s))=\tilde\jmath(u-\tilde w(s))\label{jt2}
\end{gather}
for any two given choices $j=j_i,\,w=w_i,\,(i=1,2)$. To show 
$W_\pm(J_1)=W_\pm(J_2)$ it thus suffices to prove 
$W_\pm(J)=W_\pm(\tilde J)$ for $J=J_1,\,J_2$. Since $(s,u)$ are Euclidean
coordinates in $\Omega_\pm^e$, eqs. (\ref{jt1}, \ref{jt2}) imply 
$\tilde J {\tilde J}^*J=\tilde J$ and therefore
\begin{align*}
\slim_{t\to\pm\infty}(J-\tilde J)e^{-iH_0t}E_{\Delta}(H_0)&=
\slim_{t\to\pm\infty}(1-\tilde J {\tilde J}^*)Je^{-iH_0t}E_{\Delta}(H_0)\\
&=\slim_{t\to\pm\infty}(1-\tilde J {\tilde J}^*)e^{-iHt}E_{\Delta}(H)W_\pm(J)=0
\end{align*}
by \eqref{l4a}, proving the claim.

\section{The scattering matrix at large magnetic fields}\label{proofs2}
At large magnetic fields the scattering operator acquires a universal 
behavior, 
depending only on the bending angle, but independent of
other geometric properties of the domain, as stated in 
Thm.~\ref{theorem2}. The estimate \eqref{se}, from
which the full statement of the theorem follows by density, will be 
established 
through an approximation to the evolution $e^{-iHt}\psi$ which is accurate 
at all times and not just near $t=\pm\infty$, as was the case in the 
previous section. To this end we choose an adapted \emph{gauge} 
and interpret $H$ on $\Lp{2}{\Omega}$ as a \emph{perturbation} of $H_0$ on 
$\Lp{2}{\Omega_0}$. This will require an \emph{identification} of the two 
spaces which is more accurate than \eqref{lb120}. Since these steps
are intended for the limit $\beta\to\infty$, we will assume $\beta\geq
1$ throughout this section. 

We begin with the choice of {\it gauge}, which is a deformation of Landau's.

\begin{lemma}\label{gauge}
There is a smooth vector field on $\Omega$ with $\nabla\wedge A=1$ and 
(\ref{g}, \ref{g1}) whose pull-back on $\Omega_{0}^e$ under the 
tubular map, $A_0:=(\D \mathcal{T})^t A$, is 
\nbeq\label{g2}
A_0(s,u)=-(u-\frac{u^2}{2}\kappa(s), 0).
\neeq
\end{lemma}

In the definition \eqref{so} of the scattering operator $S$ asymptotic states
are represented as states in $\dintg{}{\hilbert_T}{k}$ by means of 
$\mathcal{F}_\beta$, see \eqref{ft}. It is useful to make the band structure 
of $\widehat{H}_0$ explicit there. The range of $E_{\Delta}(H_0)$ 
then becomes isomorphic to the direct sum
\beq
E_{\Delta}(H_0)\hilbert_0\cong 
\bigoplus_{n\in \mathcal{B}}\Lp{2}{I_n,\mathrm{d}k},
\eeq
where $I_n:=E_n^{-1}[\Delta]$ is bounded and 
$\mathcal{B}:=\Set{n\in\NN}{I_n\not=\emptyset}$ is finite if $\Delta$ is as
in Thm.~\ref{theorem2}. The isomorphism is established by the unitary
\begin{gather}
\mathcal{U}:\,\bigoplus_{n\in \mathcal{B}}\Lp{2}{I_n,\mathrm{d}k}\,\rightarrow
E_{\Delta}(H_0)\hilbert_0, \qquad
\mathcal{U}=\bigoplus_{n\in \mathcal{B}}U_n\nonumber\\
\noalign{with}
U_n:\, \Lp{2}{I_n,\mathrm{d}k}\,\rightarrow E_{\Delta}(H_0)\hilbert_0,\qquad
U_nf:=\mathcal{F}_\beta(\psi_n f),\nonumber\\
\noalign{i.e.,}
(U_nf)(s)=\frac{\beta^{1/2}}{(2\pi)^{1/2}}
\intg{I_n}{e^{i\beta k s} \mathcal{D}_\beta\psi_n(k)f(k)}{k}.
\label{un}
\end{gather}
The Hamiltonian for the $n$-th band, $U_n^*H_0U_n=: h_n$, is
multiplication by $E_n(k)$. We define \emph{single band wave operators} as
\nbeq\label{sb}
\Omega_{\pm}(n):=\slim_{t\to\pm\infty} e^{iHt}JU_n e^{-ih_n t}=W_{\pm}U_n,
\neeq
and corresponding scattering operators as
\beq
\sigma_{nm}:=\Omega_+^*(n)\Omega_-(m).
\eeq
At this point \eqref{se} reduces to 
\nbeq\label{sigmaestm}
\norm[\bounded{\Lp{2}{I_m},\Lp{2}{I_n}}]{\sigma_{nm}-\delta_{nm}e^{i\phi_n(k)}}\leq C_{\Delta,\varepsilon}\beta^{-1+\varepsilon}.
\neeq

An improved \emph{identification} operator
$\tilde J: \,\Lp{2}{\Omega_0} \,\rightarrow \,\Lp{2}{\Omega}$ is
\nbeq\label{lb120m}
(\tilde J\psi)(x)=\begin{cases}
j(u-w(s))g(s,u)^{-1/4}\psi(s,u),&\text{ if } x=x(s,u)\in \Omega^{e},\\
\quad 0,& \text{ otherwise. }
\end{cases}
\neeq
It is obtained as a modification of \eqref{lb120}, where 
$g(s,u)^{1/2}=|\det\D\mathcal{T}|$ and $g\dm s\dm u$ is the Euclidean 
volume element $\dm x_1 \dm x_2$ in tubular coordinates. We take the 
parameter $w_0$ in \eqref{lb121} so that $3w_0<\inf_sw(s)$. Then $j(u-w(s))=1$
for $u<w_0$ and $\tilde J$ acts as an isometry on states supported 
near $\partial\Omega_0$, which is where we expect edge states to be 
concentrated at all times. 

The \emph{perturbation} induced by the curvature of $\partial\Omega$ on the
dynamics will be accounted for by a modification $\tilde U_n$ of $U_n$ in
\eqref{un}, resp. $\tilde{\mathcal{J}}_n:=\tilde J\tilde U_n$ of $JU_n$ in \eqref{sb}:
\begin{gather}
\tilde{U}_n:\,\Lp{2}{I_n,\mathrm{d}k}\,\rightarrow \,\hilbert_0\label{tun1}\\
(\tilde{U}_n
f)(s):=\frac{\beta^{1/2}}{(2\pi)^{1/2}}
\intg{I_n}{ e^{i(\beta k s+\phi_n(s,k))}\mathcal{D}_\beta\tilde{\psi}_n(s,k)f(k)}{k},
\label{tun2}\\
\noalign{where}
\phi_n(s,k)=-\frac{E_n^{(1)}(k)}{E'_n(k)}\sintg{-\infty}{s}{\kappa(s')}{s'}\label{addphase},\\
\tilde{\psi}_n(s,k)=
\psi_n(k)+\beta^{-1}\kappa(s)\tilde{\psi}_n^{(1)}(k),\nonumber\\
\tilde{\psi}_n^{(1)}(k)=\psi_n^{(1)}(k)-\frac{E_n^{(1)}(k)}{E'_n(k)}
\bigl((\partial_k\psi_n)(k)+
\scprd{(\partial_k\psi_n)(k)}{\psi_n(k)}\psi_n(k)\bigr),\nonumber\\
\psi_n^{(1)}(k)=-(H_0(k)-E_n(k))^{-1}(1-P_n(k))H_1(k)\psi_n(k)\label{evcorr}.
\end{gather}
It will be proved later that \eqref{tun2} yields a bounded map \eqref{tun1}. 
Here we remark that $H_1(k)\psi_n(k)$ is well-defined because $\psi_n(k)$
decays exponentially in $u$ and that $\tilde{\psi}_n(k)$ transforms as 
\eqref{ph} under a change of phase. A semiclassical 
interpretation of the above construction is in order. The evolution would 
adiabatically promote
a particle from the asymptotic state $\psi_n(k)$ at $s=-\infty$ to 
the perturbed
eigenstate $\psi_n^{[1]}(s,k)=\psi_n(k)+\beta^{-1}\kappa(s)\psi_n^{(1)}(k)$
of \eqref{sym}, if $k$ were an adiabatic invariant. It is only approximately 
so, since it changes by 
$\dm k/\dm t=\{H(s,k), k\}\approx -\beta^{-1}\dot\kappa(s)E_n^{(1)}(k)$
per unit time or, cumulatively w.r.t. arc length, by 
$\delta k(s)=-\beta^{-1}\kappa(s)E_n^{(1)}(k)/E'_n(k)$.
Therefore a more accurate state is 
$e^{i\beta^{-1}\gamma_\mathrm{B}(s,k)}\psi_n^{[1]}(s, k+\delta k(s))$, where
the phase is determined by parallel transport, see
eq.~\eqref{lb705}. For small $\beta^{-1}$ it equals $\tilde{\psi}_n(s, k)+\Order{\beta^{-2}}$.

The main intermediate result of this section is that $\tilde{J}_ne^{-ih_nt}$ is an accurate approximation of $e^{-iHt}$ at all times in the relevant energy range:
\begin{proposition}\label{mainprop}
For all $\varepsilon>0$ and $\Delta$ as in Thm.~\ref{theorem}:
\beq
\sup_{t\in\RN}\norm[\bounded{\Lp{2}{I_n},\Lp{2}{\Omega}}]{E_\Delta(H)(e^{-iHt}\tilde{\mathcal{J}}_n-\tilde{\mathcal{J}}_ne^{-ih_nt})}\leq C_{\Delta,\varepsilon}\beta^{-1+\varepsilon}.
\eeq
\end{proposition}

The implication of this result on the scattering operators $\sigma_{nm}$ can now be phrased conveniently in terms of
Isozaki-Kitada wave operators $\tilde{\Omega}_\pm(n)$:
\begin{proposition}\label{Isozaki}
The limits
\nbeq
\tilde{\Omega}_\pm(n)=\slim_{t\to\pm\infty} e^{iHt}\tilde{\mathcal{J}}_ne^{-ih_nt}\label{l12}
\neeq
exist and equal
\nbeq
\tilde{\Omega}_-(n)=\Omega_-(n),\quad \tilde{\Omega}_+(n)=\Omega_+(n)e^{i\phi_n(k)}.\label{l13}
\neeq
Moreover, for $\varepsilon>0$,
\nbeq
\norm{\tilde{\Omega}_+^{*}(n)\tilde{\Omega}_-(m)-\delta_{nm}}\leq C\beta^{-1+\varepsilon}.\label{l14}
\neeq
\end{proposition}

Since $\sigma_{nm}=e^{i\phi_n(k)}\tilde{\Omega}_+^{*}(n)\tilde{\Omega}_-(m)$,
the proof of eq.~\eqref{sigmaestm} and of Thm.~\ref{theorem2} is complete, 
except for the proofs of 
Lemma~\ref{gauge} and Props.~\ref{mainprop}, \ref{Isozaki} which we will 
give in the rest of this section.

\begin{proof} (Lemma~\ref{gauge}) We may first define $A(x)$ for 
$x\in\Omega_e$ so that \eqref{g2} holds, i.e., in terms of forms
$A=(\mathcal{T}^*)^{-1}A_0$, $A_0= -(u-\frac{u^2}{2}\kappa(s))\dm s$. We indeed have 
$\nabla\wedge A=1$ there, because 
\beq
\dm A_0=-(1-u\kappa(s))\dm u\wedge \dm s=g^{1/2}\dm s\wedge\dm u,
\eeq
and thus $\dm A=(\mathcal{T}^*)^{-1}(\dm A_0)=\dm x_1\wedge \dm x_2$, but also 
$\dm A=(\nabla\wedge A)\dm x_1\wedge \dm x_2$. We also note that \eqref{g}
holds, since $A(\gamma(s))\cdot\dot{\gamma}(s)=A_0(\partial_s)|_{u=0}=0$.
The definition of $A$ can then be
extended as follows to all of $\Omega$: Starting from any field $\tilde A$ 
with $\nabla\wedge\tilde A\equiv 1$ on $\Omega$ , there is $\chi(x)$ such that
$A=\tilde A+\nabla \chi$ on $\Omega_e$. Now it suffices to extend the scalar
function $\chi$ to $\Omega$.
\end{proof}

Some of the further analysis is conveniently phrased in terms of
pseudodifferential calculus, of which we shall need a simple version. We fix
a band $n$ with momentum interval $I_n$ and  drop the band index $n$
from all quantities throughout the remainder of this section. 
The symbols are defined on the phase space $\RN\times I\ni(s,k)$ of a 
particle on the boundary $\partial\Omega$ and take values in some Banach 
space $X$, typically $X\subset\hilbert_T$:
\nbeq\label{a2}
\mathcal{A}_2(X):=\Set{a}
{a(s,k)\in X,\,\norm[\mathcal{A}_2(X)]{a}^2:=
\intg{}{\sup_{k\in I}\norm[X]{a(s,k)}^2}{s}<\infty}.
\neeq
We abbreviate $\mathcal{A}_2\equiv\mathcal{A}_2(\hilbert_T)$. If 
$X=\dom{M}$ is the domain of some closed operator $M$ equipped with the graph 
norm $\norm[M]{\cdot}=\norm[\hilbert_T]{\cdot}+\norm[\hilbert_T]{M\cdot}$, we
just write $\mathcal{A}_2(M)\equiv\mathcal{A}_2
(\dom{M})$.

For a symbol $a\in\mathcal{A}_2(X)$, we define an operator by 
\emph{left-quantization}
\begin{gather}
\op[n]{a}:\quad \Lp{2}{I}\,\rightarrow \, \Lp{2}{\RN,X},\nonumber\\
(\op[n]{a} f)(s):=\frac{\beta^{1/2}}{\sqrt{2\pi}}\intg{I}{e^{i\beta k s}
( \mathcal{D}_\beta a)(s,k)f(k)}{k},\label{lb504}
\end{gather}
where $ \mathcal{D}_\beta$ is as in \eqref{dbeta}.
The integral is a Bochner integral on $\hilbert_T$ \cite[Thm. 1.1.4]{ABHN}. It 
exists pointwise for each $s\in\RN$ with 
$\sup_{k\in I}\norm[X]{a(s,k)}<\infty$, because $\hilbert_T$ is separable
and $\norm[1]{f}\leq |I|^{1/2}\norm[2]{f}$. Moreover, \eqref{lb504} defines a 
bounded operator 
$\op[n]{a}:\Lp{2}{I}\,\rightarrow\,\Lp{2}{\RN,X}$, because of  
\beq
\norm{\op[n]{a}}\leq
\frac{(\beta|I|)^{1/2}}{\sqrt{2\pi}}\norm[\mathcal{A}_2(X)]{a}\norm[2]{f}.
\eeq
We shall extend in two ways the class of symbols $a$ admissible in \eqref{lb504}. First, that equation defines a bounded operator  
$\Lp{2}{I}\,\rightarrow\,\Lp{2}{\RN,X}$ also if $a(s,k)$ tends to some 
asymptotes for some $a_\pm(k)$ at large $s$, in the sense that
\beq
\sintg{0}{\pm\infty}{\sup_{k\in I}\norm[X]{a(s,k)-a_\pm(k)}^2}{s}<\infty,
\qquad
\sup_{k\in I}\norm[X]{a_\pm(k)}<\infty.
\eeq
We denote such symbols by $a\in \mathcal{A}(X)$. In fact, the integral is still defined pointwise as before; in the case that $a$ is independent of $s$ the result follows by the unitarity of the Fourier transform, and in general from
$a(s,k)-\theta(s)a_+(k)-\theta(-s)a_-(k)\in\mathcal{A}_2(X)$. (Further
conditions for $\norm{\op[n]{a}}<\infty$, which we shall not need, are given 
by the Calder\'on-Vaillancourt theorem \cite{Mar}.) Second, the notation 
\eqref{lb504} shall be used also when the symbol $a(s,k)$ is 
actually a polynomial in $\beta^{-1}$, $a(s,k)=\sum_{j=0}^{\deg a}\beta^{-j}a_j(s,k)$, in which case $\norm[X]{a(s,k)}^2:=\sum_{j=0}^{\deg a}\norm[X]{a_j(s,k)}^2$. An example for both extensions is 
$a(s,k):=\tilde{\psi}(s,k)e^{i\phi(s,k)}
\in\mathcal{A}(H_0(k))$, for which 
$\op[n]{a}=\tilde{U}$. In particular \eqref{tun2} defines a bounded map, as 
claimed. Note that $\dom{H_0(k)}$, see \eqref{h0k}, is
independent of $k$.

The following propagation estimate holds:

\begin{lemma}\label{pseudoprop}
Let $a\in\mathcal{A}_2$. Then
\nbeq\label{lb50}
\sintg{-\infty}{\infty}{\norm{\op[n]{a}e^{-iht}f}^2}{t}\leq C \beta  \norm{f}^2,
\neeq
where
\beq
C=\intg{}{\sup_{k\in I}\frac{\norm{a(s,k)}^2}{E'(k)}}{s}<\infty.
\eeq
Moreover,
\nbeq
\slim_{t\to\pm\infty}\op{a}e^{-iht}=0.\label{lb50a}
\neeq
\end{lemma}

\begin{proof}
The integrand of the l.h.s of \eqref{lb50} is
\begin{multline*}%
\norm{\op[n]{a}e^{-iht}f}^2=\frac{\beta}{2\pi}\rintg{}{\rintg{I}{\rintg{I}{e^{i\beta(k_1-k_2)s}}{k_1}}{k_2}e^{-i(E(k_1)-E(k_2))t}}{s}\\
\times \scprd{a(s,k_2)}{a(s,k_1)}\bar{f}(k_2)f(k_1),
\end{multline*}
where we used that $ \mathcal{D}_\beta$ is unitary.
Formally, we may use 
\beq
\frac{1}{2\pi}\intg{}{e^{-i(E(k_1)-E(k_2))t}}{t}=\delta(E(k_1)-E(k_2))=E'(k_1)^{-1}\delta(k_1-k_2),
\eeq
because $k\mapsto E(k)$ is monotonous, so that \eqref{lb50} equals 
\nbeq \label{lb205}
\beta \rintg{}{\rintg{I}{E'(k)^{-1}
\norm{a(s,k)}^2\norm{f(k)}^2}{k}}{s},
\neeq
from which the first claim follows. More carefully, we change 
variables $k_i\mapsto E(k_i)=e_i$, 
$\mathrm{d}k_i=E'(k_i)^{-1}\mathrm{d}e_i$ 
and extend the integrand by zero for $e_i\not\in E^{-1}(I)$. Then 
\eqref{lb205} follows by Tonelli's theorem and Parseval's identity. 

Eq.~\eqref{lb50a} follows from the fact that $\op{a}e^{-iht}f$ has bounded derivative in $t$ and is square
integrable w.r.t $t$.
\end{proof}

Prop.~\ref{mainprop} states that 
$\tilde{\mathcal{J}}=\tilde{J}\tilde{U}$ approximately intertwines between the dynamics
$h$ on $\Lp{2}{I,\mathrm{d}k}$ and $H$ on $\hilbert$. Its proof will 
combine the intertwining properties of $\tilde J$ and of $\tilde U$, 
as discussed separately by the following two lemmas. 

\begin{lemma}\label{lemma3.3}
Let 
\beq
H_1:=\beta^{-1}\left(2(\beta u)D_s\kappa D_s-\frac{1}{2}(\beta
u)^2\acom{\kappa}{D_s}\right), 
\eeq
where $D_s=-i\beta^{-1}\partial_s+\beta u$. 
Then for any $1/2<\alpha\leq 1$:
\nbeq
(H\tilde J-\tilde J(H_0+H_1))\tilde{U}=
\mean{\sigma}^{-\alpha} R\,\op[n]{b},\label{l15}
\neeq
where $\norm[\mathcal{A}_2]{b}\leq C$ and $\norm[\bounded{\hilbert_0,\hilbert}]{R}\leq C_\alpha \beta^{-2}$.
\end{lemma}

\begin{lemma}\label{lemma3.5}
For any $\alpha>0$ we have:
\nbeq
(H_0+H_1)\tilde{U}-\tilde{U}h=\mean{\sigma_0}^{-\alpha}R\,\op[n]{b},\label{l16}
\neeq
where $\norm[\mathcal{A}_2]{b}\leq C$, $\norm[\bounded{\hilbert_0}]{R}\leq
C_\alpha \beta^{-2}$ and $H_1$ as in Lemma \ref{lemma3.3}.
\end{lemma}

The first lemma states that on the image of $\tilde{U}$ the Hamiltonian $H$
is a perturbation of the half-plane Hamiltonian $H_0$. 
The leading part, $H_1$, of this perturbation is formally of order 
$\beta^{-1}$, because $\beta u$ and $D_s$ are of $O(1)$ on the image 
of $\tilde{U}$. Since the tangential velocity 
$i\com{H_0}{s}=2\beta^{-1}D_s$ is of order $\beta^{-1}$, 
the size of $H_1$ is thus inversely proportional to the time 
$\sim \beta$ (in units of the inverse cyclotron frequency) required by the 
particle to traverse the bent, i.e., $\support\kappa$. 
The cumulated effect is thus of order 1, like the phase \eqref{addphase} which
by the second lemma accounts for it to leading order. Subleading contributions
occurring in either approximation are formally of order
$\beta^{-2}$. They may be integrated in time and controlled
by means of the propagation estimates in Lemmas~\ref{propagationestimate},
\ref{pseudoprop}.

\begin{proof} (Proposition \ref{mainprop})
Upon multiplication by $e^{iHt}$ the quantity to be estimated is seen to be
\nbeq
E_\Delta(H)(e^{iHt}\tilde{\mathcal{J}}e^{-iht}-\tilde{\mathcal{J}})
=i\txsintg{0}{t}{E_\Delta(H)e^{iH\tau}(H\tilde{\mathcal{J}}-\tilde{\mathcal{J}}h)e^{-ih\tau}}{\tau}\label{quanttoestm}.
\neeq
We expand
\beq
H\tilde{\mathcal{J}}-\tilde{\mathcal{J}}h=(H\tilde J-\tilde J(H_0+H_1))\tilde{U} + 
\tilde J( (H_0+H_1)\tilde{U}-\tilde{U} h),
\eeq
and insert the two terms on the r.h.s. into \eqref{quanttoestm}. We use the general fact that 
\beq
\norm{T}=\sup\Set{\abs{\scprd{\varphi_2}{T\varphi_1}}}{\varphi_i\in\hilbert_i,\,\norm{\varphi_i}=1,\,(i=1,2)}
\eeq
for operators $T:\,\hilbert_1 \,\rightarrow\,\hilbert_2$ between Hilbert
spaces, and apply the estimates (\ref{l15}, \ref{l16}) on the two 
contributions respectively. For the second term we also use 
$\tilde J\mean{\sigma_0}^{-\alpha}=\mean{\sigma}^{-\alpha}\tilde J$. 
Together with (\ref{propestm}, \ref{lb50}), we see that the two contributions are 
bounded in norm by a constant times 
$\beta^{-2}\cdot\beta^{(1+\varepsilon)/2}\cdot\beta^{1/2}=
\beta^{-1+\varepsilon/2}$.
\end{proof}

The proofs of Lemmas \ref{lemma3.3}, \ref{lemma3.5} are postponed till after that of Proposition \ref{Isozaki}.
\begin{proof} (Proposition \ref{Isozaki})
Let $F(s\in A)$ be the characteristic function of the set $A\subset \RN$.  We claim that for any $a\in\RN$ 
\nbeq
\slim_{t\to -\infty}F(s\geq -a)\tilde{U}e^{-iht}=0,\label{l8}
\neeq
and similarly for $U$ instead of $\tilde{U}$, as well as for $F(s\leq a)$ and $t\to +\infty$. It will be enough
to prove \eqref{l8} when acting on $f\in C^{\infty}_0(I)$.

We then have
\beq
(\tilde{U}e^{-iht}f)(s)
=\frac{\beta^{1/2}}{(2\pi)^{1/2}}
\intg{I}{e^{i(\beta ks-E(k)t+\phi(s,k))}\mathcal{D}_\beta\tilde\psi(s,k)f(k)}{k}
\eeq
with
\nbea
\pdiff{}{k}(\beta ks -E(k)t+\phi(s,k))&=&\beta s -E'(k)t-\diff{}{k}\left(\frac{E^{(1)}(k)}{E'(k)}\right)\sintg{-\infty}{s}{\kappa(s')}{s'}\nonumber\\
&\geq& 1+\beta\abs{s+a}+\delta\abs{t}\label{l9}
\neea
for some $\delta>0$, all $s\geq -a$ and $-t$ large enough. We may
pretend that $\tilde{\psi}(s,k)$ is replaced by $\psi(k)$, as the
difference is dealt with by \eqref{lb50a}. Since the latter amplitude is
independent of $s$, the usual non-stationary phase method
(e.g. \cite[Thm. XI.14 and Corollary]{RS3}) may be applied. We obtain (without keeping track of the dependence of constants on $\beta$)
\beq
\norm[\hilbert_T]{(\tilde{U}e^{-iht}f)(s)}\leq C_l(1+\abs{s+a}+\abs{t})^{-l},\quad (l\in\NN,\,s\geq -a),
\eeq
where we also used that $\psi(k)\in C^{\infty}(I,\hilbert_T)$. As a result,
\beq
\norm{F(s\geq -a)\tilde{U}e^{-iht}f}^2\leq C'_l(1+\abs{t})^{-2l+1},
\eeq
for $-t$ large enough, proving \eqref{l8}. As the estimate \eqref{l9} also holds with $\phi(s,k)$ omitted or replaced by $\phi(k)=\phi(s=\infty,k)$, the result applies to $U$ and $Ue^{i\phi(k)}$ as well.

We maintain that \eqref{l8} implies
\begin{gather}
\slim_{t\to -\infty}(U-\tilde{U})e^{-iht}=0,\label{l10}\\
\slim_{t\to+\infty}(Ue^{i\phi(k)}-\tilde{U})e^{-iht}=0,\label{l11}\\
\slim_{t\to\pm\infty}(J-\tilde J)Ue^{-iht}=0.\label{l11a}
\end{gather}
Indeed, if $-a<\support\kappa$, and hence $e^{i\phi(s,k)}=1$ as well as $\tilde{\psi}(s,k)=\psi(k)$ for $s< -a$, then
\beq
U-\tilde{U}=F(s\geq -a)(U-\tilde{U})
\eeq
and \eqref{l10} follows from \eqref{l8}. Eq.~\eqref{l11} is shown similarly by
using $\phi(s,k)=\phi(k)$ for $s>\support\kappa$. Eq.~\eqref{l11a}
follows
from $J-\tilde J=(J-\tilde J)F(\abs{s}\leq a)$, since $g(s,u)=1$ for 
$(s,u)\in\Omega_0^e$, $|s|\geq a$.
Now (\ref{l12}, \ref{l13}) are immediate. They follow from the existence 
of the wave operators \eqref{sb}, i.e., 
$\Omega_\pm(n)=\slim_{t\to\pm\infty} e^{iHt}JUe^{-iht}$, by means of
\eqref{l11a} and of \eqref{l10}, resp. \eqref{l11}.

Finally, we prove \eqref{l14}. Here it is necessary to introduce the
band labels again. By the intertwining property of $\tilde{\Omega}_\pm(n)$ between $H$ and $h_n$ we have 
\begin{align*}
&\scprd{g}{\tilde{\Omega}_+^{*}(n)\tilde{\Omega}_-(m)f}=\scprd{\tilde{\Omega}_+(n)g}{E_\Delta(H)\tilde{\Omega}_-(m)f}\\
&=\lim_{t\to\infty}\scprd{e^{iHt}\tilde{\mathcal{J}}_ne^{-ih_nt}g}{E_\Delta(H)e^{-iHt}\tilde{\mathcal{J}}_me^{ih_mt}f}\\
&=\lim_{t\to\infty}\scprd{\tilde{\mathcal{J}}_ne^{-ih_nt}g}{E_\Delta(H)e^{-2iHt}\tilde{\mathcal{J}}_me^{ih_mt}f}.
\end{align*}
By Proposition \ref{mainprop} this inner product equals, up to a function of $t$ bounded by $C\beta^{-1+\varepsilon}\norm{g}\norm{f}$,
the expression
\begin{gather*}
\scprd{\tilde{\mathcal{J}}_ne^{-ih_nt}g}{E_\Delta(H)\tilde{\mathcal{J}}_me^{-ih_mt}f}=\scprd{e^{iHt}\tilde{\mathcal{J}}_ne^{-ih_nt}g}{E_\Delta(H)e^{iHt}\tilde{\mathcal{J}}_me^{-ih_mt}f}\\
\stackrel{t\to+\infty}{\rightarrow}\scprd{\tilde{\Omega}_+(n)g}{\tilde{\Omega}_+(m)f}
=\scprd{e^{i\phi_n(k)}g}{\Omega_+^{*}(n)\Omega_+(m)e^{i\phi_m(k)}f}
=\delta_{nm}\scprd{g}{f},
\end{gather*}
proving \eqref{l14}. In the last line we used $\Omega_+^{*}(n)\Omega_+(m)=\delta_{nm}\mathrm{Id}_{\Lp{2}{I_m}}$. This follows from $W_+^*W_+=\mathrm{Id}_{\hilbert_0}$ and $U_n^*U_m=\delta_{nm}\mathrm{Id}_{\Lp{2}{I_m}}$.
\end{proof}

It remains to prove Lemmas \ref{lemma3.3}, \ref{lemma3.5}.

An element of pseudodifferential calculus \cite{Mar} is the symbolic 
product. We will need the product of an
operator valued symbol $h\in\mathcal{A}(\bounded{X,\hilbert_T})$ with
a vector valued one, $a\in\mathcal{A}(X)$, which in the case that $h(s,k)$ is
a polynomial in $k$ is defined as
\beq
(\moyal{h}{a})(s,k):=\sum_{l=0}^{\deg h}\frac{\beta^{-l}}{i^l l!}(\partial_k^lh)(s,k)\cdot(\partial_s^l a)(s,k),
\eeq
since the sum is then finite. In applications of this product it is understood that $a\in C_s^{\deg h}(\mathcal{A}(X))$, where $a\in C_s^{l}(\mathcal{A}(X))$ means
$\partial^j_s a\in\mathcal{A}(X)$, $0\leq j\leq l$.

\begin{proof} \begin{sloppypar}(Lemma \ref{lemma3.5})
Set $a(s,k):=\tilde{\psi}(s,k)e^{i\phi(s,k)}\in\mathcal{A}(H_0(k))$. Then by Lemma~\ref{l111},\end{sloppypar}
\nbeq\label{h0u}
H_0\tilde{U}=H_0\op[n]{a}=\op[n]{\moyal{H_0(k)}{a}},
\neeq
where $H_0(k)$ is given in \eqref{h0k}.
The operator $H_1$ may be written as
\begin{multline*}
H_1=\beta^{-1}\kappa(s)[2(\beta u)(-i\beta^{-1}\partial_s)^2+3(\beta u)^2(-i\beta^{-1}\partial_s)+(\beta u)^3]\\
-i\beta^{-2}\dot{\kappa}(s)[2(\beta u)(-i\beta^{-1}\partial_s)+3/2(\beta u)^2].
\end{multline*}
According to Lemma \ref{lb112} we have
$\moyal{k^l}{a}\in\mathcal{A}(e^{\lambda u})$, $(l=0,1,2)$, for some $\lambda>0$. Therefore, by Lemma \ref{l111}:
\beq
H_1\tilde{U}=
\beta^{-1}\op[n]{\kappa(s)(\moyal{H_1}{a})}+\beta^{-2}\op[n]{\dot{\kappa}(s)(\moyal{H_2}{a})},
\eeq
where 
\beq 
H_1(k)=2uk^2+3u^2k+u^3,\qquad
H_2(k)=-i(2uk+(3/2)u^2).
\eeq
By evaluating the expression 
\nbeq \label{moyalH0}
\moyal{H_0}{a}=\sum_{l=0}^{2}\frac{\beta^{-l}}{i^l l!}(\partial_k^lH_0)(k)\cdot(\partial_s^l a)(s,k)
\neeq
we find:
\beq
\moyal{H_0}{a}=\tilde a_{00}+\beta^{-1}a_{01}+\beta^{-2}\tilde a_{02},
\eeq
where $\tilde a_{00}=E(k)\tilde{\psi}(s,k)e^{i\phi(s,k)}$,
$a_{01}=-\kappa(s)H_1(k)\psi(k)e^{i\phi(s,k)}$ and 
$\tilde a_{02}\in\mathcal{A}_2$ 
(coefficients with a tilde may themselves contain higher order terms in 
$\beta^{-1}$). The derivation is as follows:
The r.h.s. of \eqref{moyalH0} equals
\nbeq \label{moyalH0_1}
\moyal{H_0}{a}=\left[H_0(k)\tilde{\psi}(s,k)
+\beta^{-1}(\partial_s\phi(s,k))H_0'(k)\psi(k)\right]e^{i\phi(s,k)} + O(\beta^{-2}).
\neeq
The first contribution equals
\beq
H_0(k)\tilde{\psi}(s,k)=E(k)\tilde{\psi}(s,k)+\beta^{-1}\kappa(s)\left[\frac{E^{(1)}(k)}{E'(k)}H'_0(k)-H_1(k)\right]\psi(k),
\eeq
which follows because \eqref{evcorr} provides the eigenvector at first
order, 
\beq
\bigl((H_0(k)-E(k)\bigr)\bigl(\psi(k)+\beta^{-1}\kappa(s)\psi^{(1)}(k)\bigr)\\
=\beta^{-1}\kappa(s)\bigl(E^{(1)}(k)-H_1(k)\bigr)\psi(k),
\eeq
and from taking the derivative of $(H_0(k)-E(k))\psi(k)=0$,
\beq
\bigl(H_0(k)-E(k)\bigr)(\partial_k\psi)(k)=E'(k)\psi(k)-H_0'(k)\psi(k).
\eeq
Since $\partial_s\phi=-(E^{(1)}/E')\kappa$ we see that 
the second term within the square brackets of \eqref{moyalH0_1} is canceled
inside the first one. Hence
\beq
\moyal{H_0}{a}=\left[E(k)\tilde{\psi}(s,k)-\beta^{-1}\kappa(s)H_1(k)\psi(k)\right]e^{i\phi(s,k)}+O(\beta^{-2})
\eeq
accounting for $\tilde a_{00}$ and $a_{01}$.

$\moyal{H_1}{a}$ and $\moyal{H_2}{a}$ are evaluated straightforwardly:
\begin{align*}
\beta^{-1}\kappa(s)\moyal{H_1}{a}&=\beta^{-1}a_{11}+\beta^{-2}\tilde a_{12},\\
\beta^{-2}\dot{\kappa}(s)\moyal{H_2}{a}&=\beta^{-2}\tilde a_{22},
\end{align*}
where $a_{11}(s,k)=\kappa(s)H_1(k)\psi(k)e^{i\phi(s,k)}$ and 
$\tilde a_{ij}(s,k)\in\mathcal{A}_2$. 

Collecting our expansions we get
\beq
(H_0+H_1)\tilde{U}=\op[n]{\tilde a_{00}}+\beta^{-1}\op[n]{a_{01}+a_{11}}+\beta^{-2}\op[n]{b},
\eeq
where $b\in\mathcal{A}_2$. Since $\op[n]{\tilde a_{00}}=\tilde{U}h$ and 
$a_{01}+a_{11}=0$ we conclude that
\beq
(H_0+H_1)\tilde{U}-\tilde{U}h=\beta^{-2}\op[n]{b}.
\eeq
We may extract a smooth characteristic function $\chi$ of  $\support\kappa$ from $\op[n]{b}$. Then \eqref{l16} follows with  $R=\beta^{-2}\mean{\sigma_0}^{\alpha}\chi(s)$. 
\end{proof}
Inspection of the proof shows that derivatives up to
$\ddot{\kappa}(s)$ were assumed bounded. This holds true if $\gamma\in
C^{4}$, as assumed in the Introduction.

\begin{proof} (Lemma \ref{lemma3.3})
We begin by factorizing \eqref{l15} as
\beq
(H\tilde J-\tilde J(H_0+H_1))\tilde{U}=
\mean{\sigma}^{-\alpha}\cdot Q\cdot \mean{s}^{-1}R_\lambda\mean{s}\cdot 
\mean{s}^{-1}e^{\lambda \beta u}(H_0+i)\tilde{U},
\eeq
where $\lambda>0$ is picked small, 
$R_\lambda=e^{\lambda \beta u}(H_0+i)^{-1}e^{-\lambda\beta u}$, and
\nbeq
\label{lb20}
Q=\mean{\sigma}^{\alpha}\left(H\tilde{J}-\tilde{J}(H_0+H_1)\right)
\mean{s}e^{-\lambda \beta u}.
\neeq
The claim will be established through
\begin{gather}
\label{lb46}  
\norm[\bounded{\dom{H_0},\hilbert_0}]{Q}\le C\beta^{-2},\\
\label{lb210}
\norm[\bounded{\hilbert_0,\dom{H_0}}]
{\mean{s}^{-1}R_\lambda\mean{s}}\le C,\\
\label{lb211}
\mean{s}^{-1}e^{\lambda \beta u}(H_0+i)\tilde{U}=\op[n]{b},\qquad
\norm[\mathcal{A}_2]{b}\le C,
\end{gather}
where $\norm[\mathcal{A}_2]{\cdot}$ is the norm in \eqref{a2}.

Indeed, \eqref{lb210} follows from
\beq
\mean{s}^{-1}R_\lambda\mean{s}=R_\lambda-\mean{s}^{-1}R_\lambda\com{H_0}{\mean{s}}R_\lambda
\eeq
and $R_\lambda\in\bounded{\hilbert_0,\dom{H_0}}$, $\sup_{\beta\geq 1}\norm[\bounded{\hilbert_0,\dom{H_0}}]{R_\lambda}<\infty$. 

Turning to \eqref{lb211}, we recall that by \eqref{h0u} 
$(H_0+i)\tilde{U}=\op[n]{a}$ with $a\in\mathcal{A}$ (though 
$a\notin\mathcal{A}_2$, cf.~$\tilde a_{00}$). 
For $\lambda$ small enough we have $a\in\mathcal{A}(e^{\lambda u})$
by Lemma \ref{lb112}. We conclude that 
$b=\mean{s}^{-1}e^{\lambda u}a\in\mathcal{A}_2$. 

In order to show \eqref{lb46}, we have to determine how $H\tilde{J}$ acts. 
For $\varphi\in \cinfc{\bar{\Omega}_0}$, $\varphi|_{\partial\Omega_0}=0$ 
a direct computation yields:
\beq
(H\tilde J\varphi)(x)=
\begin{cases} 
(g^{-1/4}\tilde{H}j\varphi)(s,u),& x=x(s,u)\in\Omega^{e},\\
0& \text{ otherwise,}
\end{cases}
\eeq
where $j=j(u-w(s))$ and $\tilde{H}$ is the differential operator on 
$\Omega_{0}^e$
\begin{gather}
\tilde{H}=
g^{1/4}\left(g^{-1/2}\tilde{D}_ig^{1/2}g^{ij}\tilde{D}_j\right)g^{-1/4},
\label{beltrami}\\
\tilde{D}_s=-i\beta^{-1}\partial_s+\beta u-\frac{\beta u^2}{2}\kappa(s),\qquad
\tilde{D}_u=-i\beta^{-1}\partial_u,\nonumber\\
g(s,u)=(1-u\kappa(s))^2,\qquad 
g^{ij}=\left(\begin{array}{cc} g^{-1} & 0 \\ 0 & 1\end{array} \right).\nonumber
\end{gather}
In \eqref{beltrami} summation over $i,\,j=s,\,u$ is understood. 
The expression inside the brackets is the Laplace-Beltrami 
operator in tubular coordinates on $\Omega^{e}_{0}$ associated to the 
covariant derivative $-i\beta^{-1}\nabla-\beta A$ on $\Omega^{e}$. 
Here we used Lemma \ref{gauge}.

Eq. \eqref{beltrami} has been rearranged in \cite[Thm. 3.1]{E} as 
$\tilde{H}=T+\beta^{-2}V$ with
\nbeq
\begin{gathered}\label{beltrami2}
T=\tilde{D}_sg^{-1}\tilde{D}_s-\beta^{-2}\partial_u^2,\\
V(s,u)=\frac{1}{2}g^{-3/2}\pdiff[2]{\sqrt{g}}{s}
-\frac{5}{4}g^{-2}\left(\pdiff{\sqrt{g}}{s}\right)^2
-\frac{1}{4}g^{-1}\left(\pdiff{\sqrt{g}}{u}\right)^2.
\end{gathered}
\neeq
Thus,
\nbeq\label{lb30a}
(H\tilde J\varphi)(x)=
\begin{cases} 
(g^{-1/4}(T+\beta^{-2}V)j\varphi)(s,u),& x=x(s,u)\in\Omega^{e},\\
0& \text{ otherwise.}
\end{cases}
\neeq
States of the form 
\beq
\tilde\psi(x)=
\begin{cases} 
g^{-1/4}(s,u)\psi(s,u),& x=x(s,u)\in\Omega^{e},\\
0& \text{ otherwise}
\end{cases}
\eeq
have norm 
$\|\tilde\psi\|^2=\txintg{\Omega_0^e}{|\psi(s,u)|^2}{s}\dm u\leq
\norm{\psi}^2$.
Since $V(s,u)$ is bounded on $\Omega_0^e$ and of compact
support in $s$, its contribution to $Q$ is seen to satisfy \eqref{lb46}. As 
for $T$, we write
\nbeq\label{kin}
T=(D_s g^{-1}D_s-\beta^{-2}\partial_u^2)
-\frac{\beta^{-1}}{2}\acom{(\beta u)^2\kappa g^{-1}}{D_s}
+\frac{\beta^{-2}}{4}(\beta u\kappa)^2g^{-1}.
\neeq
We next Taylor expand $g^{-1}$ to first, resp. zeroth order in $u$ in the
first two terms,
\bea
g^{-1}&=&1+2u\kappa+g^{-1}(3-2u\kappa)(u\kappa)^2,\\
&=&1+g^{-1}(2-u\kappa)(u\kappa),
\eea
and lump the remainders together with the last term of \eqref{kin}. These
three remainder contributions to \eqref{lb30a} have compact support in $s$ and
are bounded by $\beta^{-2}$ (in the graph norm of $H_0$) after multiplication 
by $e^{-\lambda\beta u}$,
as in \eqref{lb20}. They thus comply with \eqref{lb46}. The expanded terms in
\eqref{kin} are
\beq
D_s^2-\beta^{-2}\partial_u^2+\beta^{-1}\bigl(2(\beta u)D_s\kappa D_s-
\frac{1}{2}(\beta u)^2\acom{\kappa}{D_s}\bigr)=H_0+H_1.
\eeq
All this means that in proving \eqref{lb46} we may now pretend that 
$H\tilde J$ is given by \eqref{lb30a} with $T+\beta^{-2}V$ replaced by 
$H_0+H_1$. This is to be compared with 
\beq
(\tilde J(H_0+H_1)\varphi)(x)=
\begin{cases} 
(g^{-1/4}j(H_0+H_1)\varphi)(s,u),& x=x(s,u)\in\Omega^{e},\\
0& \text{ otherwise.}
\end{cases}
\eeq
The resulting commutator is computed as
\beq
\com{H_0+H_1}{j}=-i\beta^{-1}\acom{D_i}{\partial_ij}
-i\beta^{-2}
\bigl(2\beta u\acom{\kappa\partial_sj}{D_s}-(\beta u)^2\kappa\partial_sj\bigr).
\eeq
Its contribution to \eqref{lb20} is estimated by a constant times 
$e^{-\lambda\beta w_0/4}$ thanks to the choice of $w_0$ made in 
\eqref{lb120m}. Therefore \eqref{lb46} is proved.
\end{proof} 

\section{Higher order approximations: Space Adiabatic Perturbation Theory}
\label{hoa}
In this section we give an outlook on  higher order approximations
of the scattering operator. The central idea is that 
our approximation should be viewed as  an example of \emph{Space Adiabatic
  Perturbation Theory} \cite{T}.

We ultimately aim at the following generalization of Proposition \ref{Isozaki}:
\begin{proposition}\label{lb523}
For all $l\geq 1$ there exists an identification
$\tilde{\mathcal{J}}_n:\,\Lp{2}{I_n}\,\rightarrow\,\hilbert$ and a phase
function  $\phi_n^{(l-1)}(k)=\sum_{j=0}^{l-1}\beta^{-j}\phi_j(k)$ such that the limits
\beq
\tilde{\Omega}_\pm(n)=\slim_{t\to\pm\infty} e^{iHt}\tilde{\mathcal{J}}_ne^{-ih_nt}
\eeq
exist and equal
\beq
\tilde{\Omega}_-(n)=\Omega_-(n),\quad \tilde{\Omega}_+(n)=\Omega_+(n)e^{i\phi_n^{(l-1)}(k)}.
\eeq
Moreover, for $\varepsilon>0$,
\beq
\norm{\tilde{\Omega}_+^{*}(n)\tilde{\Omega}_-(m)-\delta_{nm}}\leq C\beta^{-l+\varepsilon}.
\eeq
\end{proposition}
If this proposition holds, we have
\beq
\norm{(S-S_\phi^{(l)})E_\Delta(\hat{H}_0)}\leq C_{\Delta,\varepsilon} \beta^{-l+\varepsilon},
\eeq
where $S_\phi^{(l)}=\dintg{}{\sum_ne^{i\phi_n^{(l-1)}(k)}P_n(k)}{k}$.

An immediate consequence is:
\begin{corollary}
\beq
\forall n\not=m:\quad\norm[\bounded{\Lp{2}{I_m},\Lp{2}{I_n}}]{\sigma_{nm}}=O(\beta^{-\infty}).
\eeq
\end{corollary}
This means that interband scattering is strongly suppressed at large $\beta$.

As before the improved identifications are decomposed as
$\tilde{\mathcal{J}}_n=\tilde{J}\tilde{U}_n$. The proof of
Proposition \ref{Isozaki} carries over to that of Proposition
\ref{lb523} if $\tilde{U}_n$ satisfies the following
requirements: 
\begin{align}
\slim_{t\to -\infty}(U_n-\tilde{U}_n)e^{-ih_nt}&=0,\label{lb520}\\
\slim_{t\to+\infty}(U_ne^{i\phi_n^{(l-1)}(k)}-\tilde{U}_n)e^{-ih_nt}&=0.\label{lb521}\\
\label{lb522}
\sup_{t\in\RN}\norm[\bounded{\Lp{2}{I_n},\Lp{2}{\Omega}}]{E_\Delta(H)(e^{-iHt}\tilde{\mathcal{J}}_n-\tilde{\mathcal{J}}_ne^{-ih_nt})}&\leq C_{\Delta,\varepsilon}\beta^{-l+\varepsilon}.
\end{align}

Complete proofs of the above statements will be given elsewhere
\cite{Diss}. Here we shall only present a heuristic derivation. 
 
$\tilde{J}$ intertwines more accurately between $H$ and
the $l$-th order semiclassical approximation $\hat{H}^{(l)}$ of
$T+\beta^{-2}V$, where $T$ and 
$V$ are as in \eqref{beltrami2}:
\beq
H\tilde{J} -\tilde{J}\hat{H}^{(l)}=O(\beta^{-(l+1)}).
\eeq 
Semiclassically means that $\mathcal{D}_\beta^{-1}\hat{H}^{(l)}\mathcal{D}_\beta$ can be written as the
Weyl quantization of some  symbol
$H^{(l)}(s,k)=\sum_{j=0}^{L}\beta^{-j}H_j(s,k)$, (here $L=l+2$),
where the symbols $H_j(s,k)$ don't depend on $\beta^{-1}$ anymore. 
Thus the main task is to find  $\tilde{U}_n$ such that
\beq
\hat{H}^{(l)}\tilde{U}_n-\tilde{U}_nh_n=O(\beta^{-(l+1)}).
\eeq

In the last section we invoked the adiabatic nature of the evolution
in order to motivate our construction of the approximate intertwiner $\tilde{U}_n$.
This property can be exploited more systematically by means of
\emph{Space Adiabatic  Perturbation Theory} (\SAPT{}) \cite{T}, which
allows to construct intertwiners $\tilde{U}_n$ at all orders $l$. Such 
approximations have to be sufficiently explicit of course in 
order to be of use.

\SAPT{} applies to mixed quantum systems whose Hamiltonian $\hat{H}$ is the
 quantization of some operator valued semiclassical symbol 
$H(z)\asymp\sum_{l=0}^\infty \varepsilon^lH_l(z)$
 w.r.t. some small parameter $\varepsilon$. $z\in\RN^{2d}$ is a phase
 space variable and the Hilbert space is $\Lp{2}{\RN^d,\hilbert_f}$, 
where $\hilbert_f$ is some other separable Hilbert space, called the
 space of fast degrees of freedom. In our case
 $d=1$, $z=(s,k)$, $\varepsilon=\beta^{-1}$ and
 $\hilbert_f=\hilbert_T$. 
The role of the  Hamiltonian $\hat{H}$ is played by
$\mathcal{D}_\beta^{-1}\hat{H}^{(l)}\mathcal{D}_\beta$ here.

\SAPT{} associates to each spectral band $\sigma(z)$ of the principal symbol $H_0(z)$
 that is separated by a gap from the rest of the spectrum  an \emph{effective Hamiltonian} $\hat{\mathfrak{h}}$ 
 that acts on a fixed Hilbert space $\Lp{2}{\RN^d,\mathcal{K}_r}$, where 
$\mathcal{K}_r$ can be any Hilbert space isomorphical to
 $\pi_0(z)\hilbert_f$ for any $z\in\RN^{2d}$. Here $\pi_0(z)$ is the
 spectral projector of $H_0(z)$ that corresponds to $\sigma(z)$.
 The effective Hamiltonian is the quantization of a
semiclassical symbol $\mathfrak{h}\asymp\sum_{l=0}^{\infty}\varepsilon^l\mathfrak{h}_l$.
The symbol can be computed explicitly using a recursive scheme.
In our case the spectral band $\sigma(z)$ is identified with one of
the deformed Landau levels $E_n(k)$. $\pi_0(s,k)\equiv P_n(k)$ is one
dimensional and therefore $\mathcal{K}_r\equiv \CN$. $\mathfrak{h}$ is  a $\CN$-valued symbol.

The main results of \SAPT{} imply the following statement:

\textit{The effective Hamiltonian is approximately intertwined with $\hat{H}$
by an isometry
\beq
\mathfrak{J}:\,\Lp{2}{\RN^d,\mathcal{K}_r}\,\rightarrow\,\Lp{2}{\RN^d,\hilbert_f},
\eeq
i.e.
\nbeq\label{lb501}
\hat{H}\mathfrak{J}-\mathfrak{J}\hat{\mathfrak{h}}=O(\varepsilon^{\infty}).
\neeq
Approximations to $\mathfrak{J}$ can be computed explicitly in terms
of its Weyl-symbol to any finite order in $\varepsilon$. 
}

In our context the physical meaning of this is that at any order in
$\beta^{-1}$ the motion of the particle along
the boundary is effectively one-dimensional at large $\beta$. It is
described to a very good approximation by an effective Hamiltonian on
a space with no transverse degree of freedom. The effective
Hamiltonian embodies all effects  of the transverse degree of freedom
on the longitudinal one.

Eq.\eqref{lb501} suggests that we split $\tilde{U}_n$ into
\beq
\tilde{U}_n=\mathcal{D}_\beta \mathfrak{J}^{(l)}\mathfrak{w}^{(l)},
\eeq
where $\mathfrak{J}^{(l)}$ is an approximation of $\mathfrak{J}$ up to
order $O(\beta^{-(l+1)})$ and  $\mathfrak{w}^{(l)}$ has to intertwine
$\hat{\mathfrak{h}}$ and $h_n$ up to order $O(\beta^{-(l+1)})$. 
This can be accomplished by standard WKB methods.
A formal exact intertwiner $\mathfrak{w}$ between $\hat{\mathfrak{h}}$ and $h_n$ is
constructed using generalized eigenfunctions of $\hat{\mathfrak{h}}$:
\beq
(\mathfrak{w} f)(s)=\frac{\beta^{1/2}}{\sqrt{2\pi}}\intg{I_n}{B(s,k)e^{i\beta S(s,k)}}{k},
\eeq
where formally
\begin{gather*}
\hat{\mathfrak{h}}B(s,k)e^{i\beta S(s,k)}=E_n(k)\cdot B(s,k)e^{i\beta
  S(s,k)},\\ 
\lim_{s\to-\infty}(B(s,k)e^{i\beta S(s,k)}-e^{i\beta ks})=0.
\end{gather*}
 WKB approximations 
to $B(s,k)$ and $S(s,k)$ then yield the approximate intertwiner $\mathfrak{w}^{(l)}$.

From $B(s,k)$ and $S(s,k)$ the phase function of the scattering
operator is immediate:
\beq
\phi^{(l-1)}(k)=\lim_{s\to +\infty}\phi^{(l-1)}(s,k),
\eeq
where
\nbeq\label{wkbansatz}
\phi^{(l-1)}(s,k)=-i\ln B(s,k)+\beta(S(s,k)-ks).
\neeq

The above derivation is rather formal. Neither 
did we show that
\eqref{lb520}, \eqref{lb521} hold nor is it clear from the discussion that the error terms are integrable in time
along the evolution which is necessary to prove \eqref{lb522}. The
latter seems plausible, however, because we saw in the last section
that the correction to the first order approximation of $\tilde{U}_n$ \emph{is}
integrable along the evolution. 

In fact a closer look at the technical assumptions made in \cite{T} about the
symbol $H_0(z)$ reveals that our symbol $H_0(k)$ fails to comply with
some of them.
 Apart from taking values in the unbounded operators, which causes minor technical complications,
it violates the so called \emph{gap
  condition}. This is a condition on the
growth of the symbol $H_0(k)$  with respect to $k$ relative to
the growth of the respective gaps between the deformed Landau levels.
The condition is used in the general setting of \cite{T} in
order to control the \emph{global} behavior of the various symbols  w.r.t. the phase space
variable $z$. The formal algebraic
relationships between them, which are inherently \emph{local}, are not affected. As is
pointed out in \cite[Sec.~4.5]{T} this does not mean that \SAPT{} is
not applicable. It just means that suitable
modifications to the general formalism have to be made in order to cover the 
special case at hand. 

 Based on these heuristics, 
 \SAPT{} gives us a recipe for computing the
scattering phase up to and including order $O(\beta^{-(l-1)})$:
\begin{enumerate}
\item Compute $\hat{H}^{(l)}$,
\item Compute $\mathfrak{h}$, the symbol of the effective Hamiltonian,
  that corresponds to $\hat{H}^{(l)}$ up to and including order $O(\beta^{-l})$, using the formalism of
  \cite{T}.
\item Compute the scattering phase $\phi_n^{(l-1)}(k)$ from a
  sufficiently accurate WKB approximation of the generalized
  eigenfunction of  $\hat{\mathfrak{h}}$.   
\end{enumerate} 

Following these steps we find for $\phi^{(1)}(k)=\phi_0(k)+\beta^{-1}\phi_1(k)$, dropping the band
index $n$ again,
\begin{gather*}
\mathfrak{h}_0=E(k),\quad \mathfrak{h}_1(s,k)=\kappa(s)E^{(1)}(k),\\
\mathfrak{h}_2(s,k)=E^{(1;2)}(s,k)+\kappa^2(s)E^{(2)}(k)-E'(k)(\partial_s\gamma_{RW})(s,k),
\end{gather*}
\begin{multline*}
\phi_0(k)=-\frac{E^{(1)}(k)}{E'(k)}\sintg{-\infty}{\infty}{\kappa(s')}{s'},\\
\shoveleft
\phi_1(k)=-\frac{1}{E'(k)}\sintg{-\infty}{\infty}{\left(E^{(1;2)}(s',k)+\kappa^2(s')E^{(2)}(k)\right)}{s'}\\
+\frac{1}{2}\left(\partial_k\left(\frac{E^{(1)}(k)}{E'(k)}\right)^2+\left(\frac{E^{(1)}(k)}{E'(k)}\right)^2\cdot\frac{E''(k)}{E'(k)}\right)\cdot\sintg{-\infty}{\infty}{\kappa^2(s')}{s'},\\
\end{multline*}
where
$E^{(1;2)}(s,k):=\scprd{\psi(k)}{H_2(s,k)\psi(k)}$, $H_2(s,k)$ is the
second order symbol of $\hat{H}^{(2)}$, 
$\gamma_{RW}$ as in \eqref{lb706} 
  and $E^{(2)}(k)$ is the second order correction to the eigenvalue $E(k)$ due to
  the perturbation $H_1(k)$:
\beq
E^{(2)}(k):=\sum_{m\not=n}\frac{\abs{\scprd{\psi_n(k)}{H_1(k)\psi_m(k)}}^2}{E_n-E_m}.
\eeq

The phase was computed from the WKB-ansatz \eqref{wkbansatz}
 where $S(s,k)$ has to satisfy
the Hamilton Jacobi equation up to order $\beta^{-3}$,
\nbeq\label{hamjaceq}
\mathfrak{h}(s,\partial_s S(s,k))-E(k)=O(\beta^{-3}),
\neeq
while $B(s,k)$ has to satisfy the amplitude transport equation \cite{LF}
\nbeq\label{transpeq}
\partial_s\left[B(s,k)^2\cdot\pdiff{\mathfrak{h}}{k}(s,\partial_s S)\right]=O(\beta^{-2}).
\neeq

It is possible to modify the formalism of \SAPT{} as presented in
\cite{T} and tailor it to our needs so that we can 
express $\tilde{U}_n$ at any order as an operator $\op{a}$.
The symbol $a$ is explicit enough as to  enable us to prove \eqref{lb520},
\eqref{lb521}, \eqref{lb522} rigorously by essentially the same
methods as in the last section. Moreover, the same formalism
allows for a straightforward recursive computation of the scattering
phase without reference to the concept of generalized eigenfunctions 
and their WKB approximations. 
A full account of this approach would go beyond the scope of this
paper and will be presented in \cite{Diss}.

\section{Appendix}
\subsection{Exponential decay}

\begin{lemma}\label{lb112} Let $I\subset\RN$ be a compact
  interval. For each $n\in\NN$ there exists
  $C<\infty$ such that for small $\lambda\geq 0$ and all $k\in I$:
\begin{enumerate}
\item
\nbeq\label{expdec2}
\norm[\hilbert_T]{e^{\lambda u}\psi_n(k)}\leq C.
\neeq
\item
\nbeq\label{expdec3}
\norm[\hilbert_T]{e^{\lambda u}\left(\partial_k\psi_n+
\scprd{\partial_k\psi_n}{\psi_n}\psi_n\right)}\leq C.
\neeq
\item
\beq
\norm[\hilbert_T]{e^{\lambda u}\psi_n^{(1)}(k)}\leq C,
\eeq
where $\psi_n^{(1)}(k)=-(H_0(k)-E(k))^{-1}(1-P_n(k))H_1(k)\psi_n(k)$ as in \eqref{evcorr}.
\end{enumerate}
\end{lemma}

\begin{proof} The following norms refer to $\hilbert_T$ or 
$\mathcal L(\hilbert_T)$, as appropriate. By a covering argument we may 
assume $I$ to be small as needed. 

1. Let $\Gamma\subset\rho(H(k))$, ($k\in I$), be compact. We have
\nbeq\label{expdec0}
\sup_{z\in\Gamma, k\in I}
\norm{e^{\lambda u}(H_0(k)-z)^{-1}e^{-\lambda u}}<\infty
\neeq
for small $\lambda$. In fact,
\beq
e^{\lambda u}H_0(k)e^{-\lambda u}=H_0(k)+2\lambda\partial_u
-\lambda^2
\eeq
differs from $H_0(k)$ by a relatively bounded perturbation, and is thus an
analytic family for small $\lambda$. Its resolvent,
which appears within norms in \eqref{expdec0}, is therefore bounded. 
This implies 
\nbeq\label{expdec1}
\norm{e^{\lambda u}P_n(k)e^{-\lambda u}}<\infty,
\neeq
where $\Gamma$ in 
\beq
P_n(k)=\frac{-1}{2\pi i}\ointg{\Gamma}{(H_0(k)-z)^{-1}}{z}
\eeq
is a contour encircling $E_n(k)$, ($k\in I$), counterclockwise. Since 
\eqref{expdec1} equals 
$\norm{e^{\lambda u}\psi_n(k)}\norm{e^{-\lambda u}\psi_n(k)}\geq 
c\norm{e^{\lambda u}\psi_n(k)}$ with $c>0$, eq.~\eqref{expdec2} follows.

2. We have $\partial_k P_n(k)=|\partial_k\psi_n\rangle\langle \psi_n|+
|\psi_n\rangle\langle\partial_k\psi_n|$, so that \eqref{expdec3} equals
\beq
\norm{e^{\lambda u}(\partial_k P_n)\psi_n(k)}\leq
\norm{e^{\lambda u}(\partial_k P_n)e^{-\lambda u}}
\norm{e^{\lambda u}\psi_n(k)}.
\eeq 
The claim then follows from \eqref{expdec2},
\beq
\partial_k P_n(k)=\frac{1}{2\pi i}\ointg{\Gamma}{
(H_0(k)-z)^{-1}(\partial_k H_0(k))(H_0(k)-z)^{-1}}{z},
\eeq 
as well as from \eqref{expdec0} and 
$e^{\lambda u}(\partial_k H_0)e^{-\lambda u}=\partial_k H_0$.

3. Finally, the last statement follows similarly from the representation of the
reduced resolvent
\beq
(H_0(k)-E(k))^{-1}(1-P_n(k))=
\frac{1}{2\pi i}\ointg{\Gamma}{(H_0(k)-z)^{-1}(z-E_n(k))^{-1}}{z}.
\eeq
\end{proof}

\subsection{Left-Quantization}

\begin{lemma}
\label{l111}
\begin{enumerate}
\item Let $T$ be some closed operator with $\dom{T}\subset \hilbert_T$. If $a\in\mathcal{A}(T)$ then
 $(\op[n]{a}f)(s)\in \dom{T}$ $\forall s\in\RN$ and
\beq
(1\otimes T)\,\op[n]{a}f=\op[n]{\mathcal{D}_\beta^{-1}T\mathcal{D}_\beta a}f.
\eeq

\item Let $a\in C_s^l(\mathcal{A}(X))$ for some $l\in\NN$, where 
$X\subset \hilbert_T$ with $\norm[\hilbert_T]{\cdot}\leq C \norm[X]{\cdot}$.
Then $k^l\Lmoyal a\in\mathcal{A}(X)$ and $(\op[n]{a}f)(s)$ is $l$-times differentiable in $s$ with
\beq
(-i\beta^{-1}\partial_s)^l(\op[n]{a}f)=\op[n]{\moyal{k^l}{a}}f.
\eeq

\item Let $a\in C_s^{2}(\mathcal{A}(H_0(k)))$. Then $\moyal{H_0}{a}\in\mathcal{A}(\hilbert_T)$, and $\op[n]{a}f\in\dom{H_0}$ with 
\beq
H_0\op[n]{a}=\op[n]{\moyal{H_0}{a}}.
\eeq
\end{enumerate}
\end{lemma}

\begin{proof} 1. is an immediate consequence of \cite[Proposition 1.1.7]{ABHN}.

2. The integrand $e^{i\beta ks}\mathcal{D}_\beta a(s,k)f(k)$ of $(\op{a}f)(s)$ is $l$-times  differentiable in $s$ because $a\in C^l_s(\mathcal{A}(X))$. An application of the  Leibniz rule yields
\begin{multline*}
(-i\beta^{-1}\partial_s)^{l}(e^{i\beta ks}\mathcal{D}_\beta a(s,k)f(k))=\mathcal{D}_\beta\left(\sum_{m=0}^{l}\frac{\beta^{-m}}{i^mm!}(\partial_k^mk^l)(\partial_s^m a)(s,k)\right)\\
\times e^{i\beta k s}f(k)=\mathcal{D}_\beta(\moyal{k^l}{a})(s,k)e^{i\beta ks}f(k).
\end{multline*}
Clearly $\moyal{k^l}{a}\in\mathcal{A}(X)$. In particular
$\norm[X]{\partial_s^l(e^{i\beta ks}\mathcal{D}_\beta a(s,k)f(k))}\leq C\abs{f(k)}\in\Lp{1}{I,X}$.
The claim now follows by dominated convergence.

3. We have $\dom{H_0(k)}\subset \dom{D_u^2}\cap \dom{u^2}$. Then,  by
1., we have that
\nbeq\label{lb80}
D_u^2\op[n]{a}f=\op[n]{-\partial_u^2a}f.
\neeq
Moreover 2. implies
\beq
D^2_s\op[n]{a}f=\op[n]{\moyal{k^2}{a}}f+2(\beta u)\op[n]{\moyal{k}{a}}f+(\beta u)^2\op[n]{a}f,
\eeq
where $\moyal{k^l}{a}\in\mathcal{A}(H_0(k))$, $(l=0,1,2)$.
Since  $\mathcal{A}(H_0(k))\subset\mathcal{A}(u^2)$, 1. 
implies that the r.h.s. of the last equation equals
\nbeq\label{lb81}
\op[n]{(\moyal{k^2}{a})+2u(\moyal{k}{a})+u^2a}f=\op[n]{\moyal{(k+u)^2}{a}}f.
\neeq
Combining \eqref{lb80},\eqref{lb81} we find 
\begin{multline*}
H_0\op[n]{a}f=(D_u^2+D_s^2)\op[n]{a}f=\op[n]{-\partial_u^2 a+\moyal{(k+u)^2}{a}}f\\=\op[n]{\moyal{H_0}{a}}f.
\end{multline*}
\end{proof}

\end{document}